\documentclass[smallextended]{svjour3}
\smartqed  
\usepackage{listings}
\usepackage{booktabs}
\usepackage[super]{nth}
\usepackage{rotating}
\usepackage{lmodern}
\usepackage{microtype}
\usepackage{listings}
\usepackage{xcolor}
\usepackage{amsmath}
\usepackage[most]{tcolorbox}
\usepackage{colortbl}
\usepackage{natbib}
\usepackage{hyperref}
\usepackage{graphicx}

\hypersetup{
    colorlinks=true,
    linkcolor=blue,
    urlcolor=blue,
    bookmarksdepth=4,
    citecolor=black
    }

\definecolor{codegreen}{rgb}{0,0.6,0}
\definecolor{codegray}{rgb}{0.5,0.5,0.5}
\definecolor{codepurple}{rgb}{0.58,0,0.82}
\definecolor{backcolour}{rgb}{0.95,0.95,0.92}

\journalname{Empirical Software Engineering}

\begin{document}

\title{PVAC: Package Version Activity Categorizer, Leveraging Semantic Versioning in a Heterogeneous System}
\titlerunning{PVAC: Package Version Activity Categorizer}

\author{Shane K. Panter \and
        Lucas S. Hindman \and
        Nasir U. Eisty
}

\institute{
    S. K. Panter [0009-0009-1852-3655],
    L. S. Hindman [0009-0004-8949-9342]
    \at
    Computer Science,
    Boise State University, USA \\
    \email{\{shanepanter, lukehindman\}@boisestate.edu}
    \and
    N. U. Eisty [0000-0001-5228-4664] \at
    EECS, University of Tennessee Knoxville, USA \\
    \email{neisty@utk.edu}
}

\date{Received: Nov 6, 2024 / Accepted: date}

\newcommand\revised[1]{{\color{black}#1}}

\newcommand\RQone{How well does PVAC categorize versioning schemes in a
heterogeneous system, and to what extent is semantic versioning followed?}

\newcommand\RQtwo{How well does the Version Number Delta (VND) component detect
activity in an ecosystem compared to established methods?}

\newcommand\RQthree{How well does PVAC categorize the activity of an individual package within a heterogeneous system?}

\newcommand\datasetSize{22,535}
\newcommand\tprate{21,027}

\maketitle

\begin{abstract}
\textbf{Context:}
Modern open-source software ecosystems, such as those managed by GNU/Linux
distributions, are composed of numerous packages developed independently by
diverse communities. These ecosystems employ package management tools to
facilitate software installation and dependency resolution. However, these tools
lack robust mechanisms for systematically evaluating the development activity
and versioning dynamics within their heterogeneous software environments.
\textbf{Objective:}
This research aims to introduce a systematic method and a prototype tool for
assessing version activity within heterogeneous package manager ecosystems,
enabling quantitative analysis of software package updates.
\textbf{Method:}
We developed a \underline{\textbf{P}}ackage \underline{\textbf{V}}ersion
\underline{\textbf{A}}ctivity \underline{\textbf{C}}ategorizer (PVAC) that consists of three components.
The Version Categorizer (VC), which  categorizes diverse semantic version numbers, a Version Number Delta (VND) 
component, \revised{which calculates a numeric score representing the aggregated semantic version changes across packages at the ecosystem level,} and finally, an Activity Categorizer (AC)  that categorizes the
activity of individual packages within that ecosystem. PVAC utilizes tailored regular
expressions to parse semantic versioning details (epoch, major, minor, and patch
versions) from diverse package version strings, enabling consistent
categorization and quantitative scoring of version changes.
\textbf{Results:}
PVAC was empirically evaluated using a dataset of \datasetSize{} packages drawn from
recent releases of Debian and Ubuntu GNU/Linux distributions. Our findings
demonstrate PVAC's effectiveness for accurately categorizing versioning schemes
and quantitatively measuring version activity across releases. We provide
empirical evidence confirming that semantic versioning, including adapted
variations, is predominantly employed across these ecosystems.
\textbf{Conclusions:}
PVAC represents an effective solution for systematically assessing and
monitoring the software package version activity within heterogeneous ecosystems. By
providing clear metrics for software activity at both the ecosystem \revised{and individual package levels, PVAC helps software maintainers and researchers precisely identify packages that require updates or security remediation,} thereby reducing
potential security risks, technical debt, and technical lag.


\end{abstract}
\section{Introduction}
\label{sec:intro}

Modern software is built on a foundation of many software packages.
Package managers provided as part of the operating system are heterogeneous, in contrast to a homogeneous system such as the Node Package Manager
(NPM)\footnote{https://www.npmjs.com/}. A fundamental distinction between the two lies in their handling of
software diversity: homogeneous package managers are comprised of software
packages that conform to a standardized versioning scheme, are authored in the
same language, and typically use the same build
system~\citep{pinckneyLargeScaleAnalysis2023a}. This uniformity simplifies
dependency resolution, update tracking, and version compatibility management. In
contrast, heterogeneous package managers~\citep{nussbaumUltimateDebianDatabase2010}, such as Debian's Advanced Packaging Tool\footnote{https://www.debian.org/doc/manuals/debian-faq/pkgtools.en.html} (apt), manage software from diverse sources, each
potentially following distinct programming languages, build systems, and
versioning strategies. This inherent diversity poses challenges in uniformly
assessing software activity at the ecosystem level, necessitating more
sophisticated tools and methodologies for systematic evaluation.

In this paper, we present the Package Version Activity Categorizer
(PVAC), a method and a prototype tool that leverages semantic versioning
information to systematically assess ecosystem activity within a heterogeneous package
manager system. At its core, PVAC can categorize a diverse set of version
numbers with the Version Categorizer (VC) component, calculate a single number to
\revised{measure the technical lag} of the ecosystem with the Version Number Delta (VND)
component, and drill down to individual packages using the Activity Categorizer
(AC) component for fine-grained analysis. By categorizing the semantic version number with the AC
component and quantifying version activity with the VND and AC components, PVAC can
provide clear and consistent assessments of a software ecosystem that includes
all the software required to run a complete system, not just packages that are
directly linked to an application. \revised{This systematic measurement helps maintainers identify potentially outdated or insecure packages that require detailed security analysis or updates}, thereby reducing security
vulnerabilities and technical lag in software ecosystems.

Systems built upon the foundation of a distribution such as Debian provide
flexibility and have been used for smartphones, servers, and desktops. However,
issues like the HeartBleed bug in OpenSSL
(CVE-2014-0160) \citep{redhatCVE20140160OpenSSLHeartbleed2014}, ShellShock in
BASH (CVE-2014-6271) \citep{debiangnu/linuxCVE20146271GNUBourneAgain2014}, and
more recently the out-of-bounds memory access in GLIBC qsort()
(CVE-2023-6246) \citep{redhatCVE20236246GlibcHeapbased2024} all occur in
packages that are necessary for a system to run, but may not be directly linked
to an application, thus obscuring potential vulnerabilities  in the software
supply chain. Tools and methods capable of systematically evaluating the entire
software ecosystem \revised{are} essential for identifying issues to reduce technical
lag~\citep{gonzalez-barahonaTechnicalLagSoftware2017} and improve the security
and stability of a system.

The main contributions of this paper are as follows:
\begin{itemize}
    \item We demonstrate that it is possible to categorize semantic
    version information in a heterogeneous software system.
    \item We present a novel method \revised{to measure the technical lag of a software ecosystem.}
    \item We demonstrate the ability to view the activity of individual packages for fine-grained analysis.
    \item We have constructed and shared a dataset \revised{that the research community can use} for future work and extensions.
\end{itemize}

The remainder of this paper is organized as follows:
Section~\ref{sec:background} provides the information to
frame the research questions. Section~\ref{sec:questions} describes our research
questions and is the \revised{paper's primary focus}. We then follow with a related work discussion in
Section~\ref{sec:related}. Section~\ref{sec:methodology} describes our
methodology\revised{,} while Section~\ref{sec:results} presents the
results. Section~\ref{sec:discussion} presents our discussion of the results, and
we explore the threats to validity in Section~\ref{sec:threats}. Finally,
Section~\ref{sec:conclusion} concludes the paper and discusses future work.
\section{Background and Motivation}
\label{sec:background}

In the early days of software engineering, there was no well-defined concept
regarding external dependencies and dependency management; companies would
implement entire systems from scratch\footnote{The invention of Unix.
https://www.bell-labs.com/institute/blog/invention-unix/}. As software development
matured and open-source software (OSS) became more prevalent~\citep{miller_we_2023, goggins_open_2021,
guizaniRulesEngagementWhy2023}, it became clear that reusing code was a more
efficient way to develop software. Thus, package managers emerged
as a method to manage these reusable components with minimal effort.
Package managers are \revised{ubiquitous in modern software development and
come in homogeneous and heterogeneous flavors.} Our goal is to 
assess the activity of software managed by heterogeneous systems using 
only the semantic versioning information.

An early example of a heterogeneous package manager is from the Debian
project, which introduced the \texttt{apt} package manager~\citep{nguyenLifeDeathSoftware2012}
in 1998. \texttt{apt} was designed to simplify software maintenance within Debian-based
distributions by automating the maintenance process. Unlike homogeneous package managers,
such as \texttt{npm}, which are typically built around a single programming language,
directly linked to a final application, and updated independently, heterogeneous package
managers such as \texttt{apt} provide critical packages such as \revised{\texttt{libc}\footnote{https://www.gnu.org/software/libc/}, language
runtimes such as \texttt{Node.js}\footnote{https://nodejs.org/en}, and system utilities such as \texttt{bash} that are necessary}
for the system to run and \revised{are} installed outside of a homogeneous package manager.
Practitioners  must take into account  the activity of the entire ecosystem, even if a
direct link is not apparent, to reduce security vulnerabilities and technical lag~\citep{gonzalez-barahonaTechnicalLagSoftware2017}.

The following sections provide the necessary background to understand
our work. We first discuss semantic versioning in
Section~\ref{sec:semantic-versioning} as it is the core focus of PVAC. We then
provide an overview of the health of OSS in Section~\ref{sec:health-oss-background} to
highlight efforts in the community on software health, Section~\ref{sec:package-manager} provides a
brief overview of package managers and finally, Section~\ref{sec:technical-lag} discusses technical lag.

\subsection{Semantic Versioning}
\label{sec:semantic-versioning}

Semantic versioning~\citep{preston-werner_semantic_nodate}, commonly known as
\texttt{semver}, is a powerful method helping developers understand the
impact of upgrading a package. It is a method that conveys the nature of
modifications of a software package in the version number itself. The core
principle of semantic versioning is to communicate the impact of changes on
compatibility and functionality to users and developers. By examining the
semantic version information of the packages, we can \revised{assess} the
system as a whole. If we look at the package manager as an
ecosystem\footnote{https://chaoss.community/kbtopic/ecosystem/}, we can
leverage the semantic versioning information to assist practitioners in
assessing entire ecosystems, not just individual packages.

In software development, there is a common term called ``Dependency
Hell''\footnote{This is a humorous homage to the term
\href{https://en.wikipedia.org/wiki/DLL_hell}{DLL Hell} \revised{which} originated \revised{in}
early Microsoft systems.}. This term refers to a problematic situation
arising when a software package has complex or nested dependencies, where each
dependency may depend on further packages, some of which may not be
apparent. Such intricate dependency chains can lead to software failing to run
due to unmet dependencies, making identifying, resolving, and
installing the correct dependencies particularly challenging or, in some cases,
virtually impossible. While the ``Dependency Hell'' problem has not been
completely solved, methods such as semantic versioning have been created to help
mitigate the problem. The core principles of semantic versioning are as follows:

\begin{itemize}
    \item \textbf{Epoch} - \revised{An optional leading integer in version strings indicating a significant, non-backward-compatible update. The epoch number precedes semantic version fields (major, minor, patch) and signals substantial changes beyond typical semantic versioning increments.}
    \item \textbf{Major} - This number is incremented when developers make breaking changes
    to the Application Programming Interface (API)
    \item \textbf{Minor} - This number is incremented when \revised{adding} functionality
    in a backward-compatible manner
    \item \textbf{Patch} - The patch version number is incremented when \revised{making}
        backward-compatible bug fixes.
\end{itemize}

\revised{For} example, consider a package that has a version of \texttt{3.1.0}
when it is incorporated into a project. A consumer of this package could safely
specify a dependency greater than or equal to \texttt{3.1.0} but less than
\texttt{4.0.0}. When bug fixes are applied, the developers increment the
patch version to \texttt{3.1.1}, indicating that it is still compatible.  If new
backwards-compatible features are added, the minor version may be incremented to
\texttt{3.2.1}. Only when a substantial modification renders the library
incompatible with preceding versions would the major version number be
incremented to \texttt{4.0.0}, indicating a breaking change. This system allows
developers to quickly assess the impact of upgrading a package based on the
version number alone~\citep{preston-werner_semantic_nodate}.

\subsection{Health of OSS}
\label{sec:health-oss-background}

The Linux Foundation has significant motivation to ensure that OSS stays healthy
and \revised{has} founded a working group referred to as Community Health Analytics in Open Source \revised{Software (CHAOSS)\footnote{https://chaoss.community/about-chaoss/}} in
2017. CHAOSS is dedicated to
creating metrics, methodologies, and software to help stakeholders understand
and evaluate the health of software projects. CHAOSS defines metrics across
various dimensions to provide a comprehensive view of a project. A
project can select the metrics that best capture the required facets to communicate the status and health of the project. \revised{Not all metrics are relevant to every project, and the CHAOSS project encourages
selecting metrics most applicable to a specific project or
organization.}

Given our goal of assessing the activity of packages installed with heterogeneous package managers, we
are particularly interested in the ecosystem
family\footnote{https://chaoss.community/kbtopic/ecosystem/} of metrics, which
provides a set of metrics to evaluate the health of the entire system rather
than a single individual project. For example, the Libyears metric is based on a  paper~\citep{cox_measuring_2015} that laid the foundations for our
research. The Libyears metric focused only on Java systems that manage their
dependencies through Maven\footnote{https://maven.apache.org/} and primarily
focused on the release dates of the package, not the semantic version number. Our
goal is to extend this concept to heterogeneous package managers with no unified
semantic versioning scheme, which presents a unique challenge to solve.

\subsection{Package Managers}
\label{sec:package-manager}

Package managers are tools that install, remove, upgrade, and configure packages in an
automated manner. These tools do not have to
be specific to a particular language, such as \texttt{npm}. They can be used independently to install source code,
executables, libraries, or plain text. \revised{Package managers can be straightforward
and only install software, or they can be very complex and manage dependencies
and versions automatically. We are interested in package managers
installing software from diverse projects and sources for this paper}.

For a package manager such as \texttt{apt}, maintaining software packages involves several roles working
together across different layers of the software supply chain. At the top are 
upstream developers, who create and evolve the source code of a
software project. \revised{Distribution package maintainers adapt the upstream source for
inclusion in a specific distribution, like Debian or Ubuntu. Finally, distribution maintainers coordinate the integration of
all packages into a cohesive, stable release~\citep{nguyenLifeDeathSoftware2012}.}

While package managers excel at installing software and managing
dependencies, they do not inherently provide mechanisms to evaluate the overall
version activity within the software ecosystem. One significant consequence of
insufficient evaluation and updating of package versions is known in the
literature as technical lag~\citep{gonzalez-barahonaTechnicalLagSoftware2017},
which represents how outdated or behind the packages within a system are
relative to the latest available versions. Thus, a methodology and tool support
for managing dependencies is essential for developers to ensure their
software is up-to-date and secure.

\subsection{Technical Lag}
\label{sec:technical-lag}

\citet{gonzalez-barahonaTechnicalLagSoftware2017} coined the term technical lag
to describe the difference between the version of a package used by a project
and the latest version of the package; it is a theoretical model that measures how outdated a software
system or its dependencies are. \revised{The
package users must first define the ``gold standard'' to measure technical lag.} What qualifies as
a ``gold standard'' is different for every system. Once you have the standard,
you must define distance metrics or lag functions to compute the lag. Lag typically
arises due to the complexity and resource costs associated with regularly
updating dependencies, even though timely updates are critical for software
quality, security, and reliability~\citep{stringerTechnicalLagDependencies2020}.

However, assessing and managing technical lag is considerably more challenging at the ecosystem
level due to the diverse
independent versioning schemes, as with
operating system distributions. \revised{The variability and complexity of versioning
schemes can obscure the system's activity, making it difficult to track the current state systematically.} Consequently, maintainers and
developers lack effective mechanisms for clearly identifying stale or inactive
packages, directly amplifying risks associated with technical lag.

\section{Research Questions}\label{sec:questions}

This section describes the research questions that drive the study.
Our work aims \revised{to detect} activity within a
heterogeneous package manager system at both the ecosystem level and individual package level by leveraging the information of each
package in the context of the system as a whole. Many studies have been
conducted on activity within homogeneous package manager
systems~\citep{cox_measuring_2015,dannUPCYSafelyUpdating2023, decanWhatPackageDependencies2021,dietrichDependencyVersioningWild2019}, but to
our knowledge, no study has been conducted on semantic versioning activity in
heterogeneous systems. Therefore, we propose the following research questions
(RQs):

\begin{description}
\label{research_questions}

\item[\textbf{RQ1:}]\textbf{\RQone}

Categorizing versioning schemes refers to systematically identifying
and categorizing package version numbering conventions. Each software package may employ varying versioning schemes, 
\revised{from strictly following semantic versioning, adhering to slightly modified
semantic structures (e.g., including epochs), or employing custom
distribution native formats. PVAC was explicitly} designed to recognize and
categorize these diverse versioning formats automatically.

\item[\textbf{RQ2:}]\textbf{\RQtwo}

To track the activity of packages, it is helpful to be able to assign a
score to the system as a whole. A single score can be communicated and
integrated into a continuous integration and deployment (CI/CD) pipeline.
PVAC introduces a method to determine a single score using only the semantic
versioning information. In this research question, we determine the
optimum VND weights through empirical analysis and compare the results to a
more traditional approach using the release date.

\item[\textbf{RQ3:}]\textbf{\RQthree}

In addition to a total system view, we look at individual packages to determine
the activity level. This allows the user to identify potential problematic
software for a deeper analysis.

\end{description}

\section{Related Work}
\label{sec:related}

Our analysis of the related work found three related threads of research: the
health of OSS, semantic versioning, and technical lag. We will discuss each of
these threads in
Sections \ref{sec:health-oss-related}, \ref{sec:semantic-versioning-related},
and \ref{sec:technical-lag-related}.

While \revised{the} existing literature thoroughly explores package management in homogeneous
ecosystems such as \texttt{npm}, our study specifically addresses unique challenges
posed by heterogeneous ecosystems like Debian's \texttt{apt}, \revised{which integrates} diverse
versioning schemes and multiple upstream projects. This distinction is crucial as it
highlights the need for a tailored approach to assess and categorize package activity
across a varied landscape of software projects.

\subsection{Health of OSS}
\label{sec:health-oss-related}

Metrics to communicate and keep OSS projects healthy \revised{are} an open research topic
that has been studied by many researchers~\citep{tanHowCommunicateWhen2019,
linaker_how_2022, dijkersExploringEffectSoftware2018, goggins_open_2021, guizaniRulesEngagementWhy2023}. These efforts
will continue to grow due \revised{mainly} to government requirements for creating secure and
reliable software. \citet{linaker_how_2022} conducted a snowball literature review
of publications on the health of OSS projects and found 107 health characteristics
divided among 15 themes that can be used to characterize the health of OSS projects.
They also confirmed the importance of not analyzing the OSS project in isolation.
We must consider its dependencies and ties to other projects to get a complete view
of its health.

Package freshness is an \revised{essential} aspect of the health of an OSS project.
Freshness relates to how up-to-date the software is, with stale software being
more likely to have security vulnerabilities~\citep{cox_measuring_2015}.
\citet{legayPackageFreshnessLinux2020, legay_quantitative_2021} studied the
freshness of GNU/Linux users to determine their perception of package freshness
in various distributions. They found that users perceive distributions such as Debian
and CentOS as stable, meaning they take several months to get new
versions. In contrast, distributions such as Arch and OpenSUSE Tumbleweed are
perceived as fresh, meaning that the software is updated as soon as it is
available.

Finally, if a project is not healthy, it may be abandoned. Researchers have
studied the abandonment~\citep{avelinoAbandonmentSurvivalOpen2019}  of OSS
projects to determine how to prevent it, how to deal with it when it
occurs~\citep{miller_we_2023}, and how to predict it~\citep{li_ossara_2022}.
Metrics such as the Truck Factor and the OSS Abandonment Risk Assessment
(OSSARA)~\citep{li_ossara_2022} model \revised{have} been defined and used for OSS
projects to determine how many developers would need to leave a project before
it would fail.

\subsection{Semantic Versioning}
\label{sec:semantic-versioning-related}

Semantic versioning has been studied in the context of package managers and
dependencies~\citep{dietrichDependencyVersioningWild2019,decanWhatPackageDependencies2021}
to understand how well projects follow semantic versioning and how to improve the adoption of
semantic versioning. \revised{Since}
semantic versioning is just a policy, and there is no enforcement mechanism,
there have been some high-profile cases of breaking changes introduced in minor
and patch versions violating the semantic versioning policy~\citep{jayasuriyaExtendedStudySyntactic2025c}.

In addition to researching how well projects follow the semantic version policy,
research into how to detect breaking changes in package
managers~\citep{kongBetterComprehensionBreaking2024}  has been conducted, \revised{and} the reasons why projects do not follow the semantic versioning policy have
been studied. \citet{jayasuriyaUnderstandingBreakingChanges2023} found that
almost half of the detected clients impacting breaking changes violate the
semantic versioning scheme by introducing breaking changes in non-major updates.

\subsection{Technical Lag}
\label{sec:technical-lag-related}

Technical lag represents a significant concern in software development and maintenance, 
occurring when software systems or their dependencies are not promptly updated to their
most recent or stable versions. This phenomenon typically emerges from the trade-offs
development teams must manage, balancing the immediate cost and risk of updating
against the longer-term consequences of using outdated software, such as increased
vulnerability to security threats, reduced compatibility, and potential system 
instability~\citep{stringerTechnicalLagDependencies2020, 
gonzalez-barahonaCharacterizingOutdatenessTechnical2020}.

Previous research highlights the widespread occurrence and impact of technical lag, 
notably within homogeneous package ecosystems such as \texttt{npm}, Maven, and PyPI~\citep{
decanEvolutionTechnicalLag2018, decanWhatPackageDependencies2021, 
lamPuttingSemanticsSemantic2020, zeroualiEmpiricalAnalysisTechnical2018, zeroualiFormalFrameworkMeasuring2019,
zeroualiMultidimensionalAnalysisTechnical2021}. These studies underscore that
technical lag can lead to severe security vulnerabilities, as outdated software packages
may lack critical patches or updates necessary for mitigating newly discovered exploits.
Moreover, compatibility issues caused by technical lag can escalate integration difficulties,
forcing developers into costly refactoring or adaptation efforts to ensure compatibility with
more recent versions or other software dependencies.

Recent advancements include automated solutions to detect and minimize technical 
lag through dependency management tools, version alerts, and automated patching
mechanisms~\citep{dannUPCYSafelyUpdating2023}. However, these tools predominantly cater
to homogeneous environments and rely on clear semantic versioning and consistent 
package management practices. In heterogeneous environments, the diversity in versioning
schemes and update frequencies demands a more sophisticated, adaptable approach, \revised{which is} precisely
what our research aims to address.

\subsection{Summary of Related Work}

Prior studies primarily explored software ecosystem health, semantic
versioning, and technical lag within homogeneous software systems. Our research
extends these findings by developing methods and a prototype tool for
heterogeneous package managers. This explicit focus on heterogeneity addresses a previously
overlooked challenge, enhancing practical applicability and theoretical
understanding of software ecosystems' dynamics.
\section{Methodology}
\label{sec:methodology}

\begin{figure}[htb]
    \centering
     \includegraphics{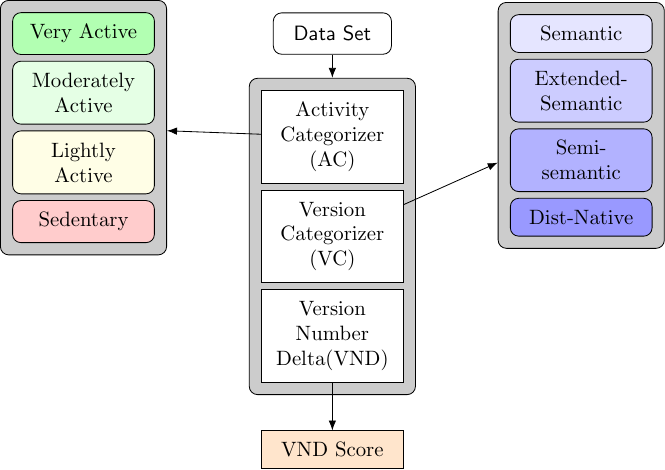}
     \caption{\revised{High-level} architecture of PVAC. The architecture consists of three main components
        shown in the center: the Activity Categorizer (AC), the Version Categorizer (VC),
        and the Version Number Delta (VND). The left side shows the possible
        activity levels for the packages in the dataset. The right side shows
        the possible version categories for the packages in the dataset.}
     \label{fig:pvac-architecture}
 \end{figure}

In this section, we present our methodology and dataset construction. The high level
architecture is depicted in Fig.~\ref{fig:pvac-architecture} and is composed
of three main components: the \textit{Version Categorizer} which is described in
Section~\ref{sec:version_categorizer} and evaluated in Section~\ref{sec:rq1}, the
\textit{Version Number Delta} component which is described in
Section~\ref{sec:version_number_delta} and evaluated in Section~\ref{sec:rq2},
and the \textit{Activity Categorizer} which is described in
Section~\ref{sec:activity_categorizer} and evaluated in Section~\ref{sec:rq3}.
Finally, we describe our dataset and how it was constructed in Section~\ref{sec:dataset}.

\subsection{Version Categorizer (VC)}
\label{sec:version_categorizer}

\begin{figure}[t]
    \centering
    \includegraphics[scale=0.8]{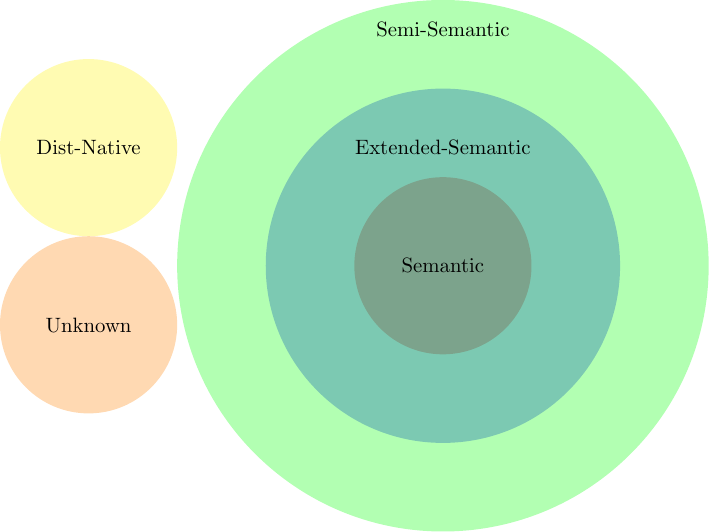}
    \caption{Euler diagram showing the version categories and \revised{their relation}. The Semantic
    category is the most common for packages that follow the official
    semantic versioning specification. The Extended-Semantic category is used for
    packages that prepend an epoch value to the version string. The
    Semi-Semantic category allows leading zeros and makes the patch field optional. The
    Dist-Native category is used for packages that are specific to a distribution.
    The Unknown category is used
    for packages with version strings that do not match any known
    category.}
    \label{fig:pvac-semantic-categories}
\end{figure}

The Version Categorizer component categorizes the
version strings of each package in a heterogeneous package management system.
The contribution of the Version Categorizer is the ability to
categorize version strings from various versioning schemes into one of four
categories: Semantic, Extended-Semantic, Semi-Semantic, and  Dist-Native. The
Semantic class is a well-known standard~\citep{preston-werner_semantic_nodate}
that is widely used in the software industry. The Extended-Semantic and
Semi-Semantic categories are additions that are described in
Sections~\ref{sec:extsemantic} and \ref{sec:semisemantic}. They are supersets of
the official semantic version standard. The Dist-Native is described in Section~\ref{sec:distnative} and is used for version strings specific
to a distribution. Unknown is used for version strings that do not conform
to the other categories and need new regular expressions or custom
code to categorize them. We visually show the relationship between the
categories as an Euler diagram in Fig.~\ref{fig:pvac-semantic-categories}.

The Version Categorizer uses a carefully validated set of regular expressions,
which were iteratively developed by analyzing and manually labeling a random
sample of packages from our dataset. PVAC takes advantage of \revised{several} common
patterns that can easily be combined into a single regular expression to reduce
the total number of expressions necessary. Listing~\ref{lst:regular_expression}
shows a sample of the regular expressions used to categorize the version strings\revised{, and
Section~\ref{sec:alternative-approaches} explores alternative approaches that could
be used.}

\vspace{.5cm} 

\begin{minipage}{.95\linewidth} 
\begin{lstlisting}[
    caption = {Example regular expressions that will categorize a package that uses
              semantic versioning in a heterogeneous package system like apt.},
    captionpos=b,
    label=lst:regular_expression,
    language=Python
    ]
    # Official Semantic
    {"pattern":re.compile(r"^(?P<major>0|[1-9]\d*)\.(?P<minor>0|[1-9]\d*)\.(?P<patch>0|[1-9]\d*)(?:-(?P<prerelease>(?:0|[1-9]\d*|\d*[a-zA-Z-][0-9a-zA-Z-]*)(?:\.(?:0|[1-9]\d*|\d*[a-zA-Z-][0-9a-zA-Z-]*))*))?(?:\+(?P<buildmetadata>[0-9a-zA-Z-]+(?:\.[0-9a-zA-Z-]+)*))?$"), "class_group":"Semantic"},
    # ExtendedSemantic: Match epoch prepended to version string based upon official apt versioning policy
    {"pattern":re.compile(r"^((?P<epoch>0|[1-9]\d*):)?(?P<major>0|[1-9]\d*)\.(?P<minor>0|[1-9]\d*)\.(?P<patch>0|[1-9]\d*)(?:-(?P<prerelease>(?:0|[1-9]\d*|\d*[a-zA-Z-][0-9a-zA-Z-]*)(?:\.(?:0|[1-9]\d*|\d*[a-zA-Z-][0-9a-zA-Z-]*))*))?(?:\+(?P<buildmetadata>[0-9a-zA-Z-]+(?:\.[0-9a-zA-Z-]+)*))?$"), "class_group":"ExtSemantic"},
    # SemiSemantic: Allow version numbers to start with 0; Make the patch field optional and separated by either a . or a lower-case p
    {"pattern":re.compile(r"^((?P<epoch>0|[1-9]\d*):)?(?P<major>[0-9]\d*)\.(?P<minor>[0-9]\d*)((\.|p|pl)(?P<patch>[0-9]\d*))?(?:-(?P<prerelease>(?:[0-9]\d*|\d*[a-zA-Z-][0-9a-zA-Z-]*)(?:\.(?:[0-9]\d*|\d*[a-zA-Z-][0-9a-zA-Z-]*))*))?(?:\+(?P<buildmetadata>[0-9a-zA-Z-]+(?:\.[0-9a-zA-Z-]+)*))?$"), "class_group":"SemiSemantic"},

\end{lstlisting}
\end{minipage}

\subsubsection{Extended-Semantic Category}\label{sec:extsemantic}

The Debian policy documentation defines the requirements for distributions that
use the apt package management system. This policy describes an optional
\textbf{epoch} value that can be prepended and an optional
\textbf{debian\_revision} that can be appended, encapsulating the original
upstream version information. \revised{According to} the Debian versioning policy, ``[Epoch] is a
single (generally small) unsigned integer. It may be omitted, in which case zero
is assumed. When comparing two version numbers, first the epoch of each is
compared, then the upstream\_version if epoch is equal, and then
debian\_revision if upstream\_version is also
equal.''~\citep{jacksonDebianPolicyManual2024}

The Extended-Semantic category checks for the prepended epoch value at the
head of an otherwise semantic version string. Based upon the Debian versioning
policy, the Semantic category would also be considered Extended-Semantic,
with the epoch value of 0 omitted. The categorizer always applies the most specific
category, so if a version string matches both the Semantic and
Extended-Semantic regular expressions, it will be categorized as Semantic.

\subsubsection{Semi-Semantic Category}
\label{sec:semisemantic}

The Semi-Semantic category removes additional constraints \revised{to categorize as many packages as possible without manual intervention}.
We define the Semi-Semantic category as follows:

\begin{enumerate}
    \item Allow version numbers with leading zeros, such as 001.
    \item Make the patch field optional. Packages without patch fields are
    assigned a patch value of 0.
    \item Recognize patch fields separated from the minor field by
    either a `.' or a lowercase `p' or `pl'.
\end{enumerate}

\subsubsection{Dist-Native Category}
\label{sec:distnative}

The Dist-Native category is used for packages specific to a
distribution and don't have an upstream source. In the case of the
Debian distribution, the mentor's
FAQ\footnote{https://wiki.debian.org/DebianMentorsFaq} describes the two types
of packages in a Debian system and their derivatives: native and non-native. PVAC
categorizes native packages as Dist-Native, while non-native packages are
categorized as Semantic, Extended-Semantic, or Semi-Semantic. For
example, the package
\texttt{bedtools} (Table~\ref{tab:rq1-sample-dataset}) has been modified to conform to Debian policies.

The majority of packages in the Debian distribution are non-native, meaning 3rd
parties maintain them and may be used by a variety of Linux distributions.
Keeping packages labeled as Dist-Native provides an interesting view of
distribution-maintained packages vs external packages.

\subsection{Version Number Delta (VND)}
\label{sec:version_number_delta}

The CHAOSS project defines Libyears as ``[w]hat is the age of the project's
dependencies compared to the current stable
releases''.~\footnote{https://chaoss.community/kb/metric-libyears/}
\citet{cox_measuring_2015} introduced this metric for evaluating OSS packages
and described several methods of calculating the final score. One
method describes using the semantic version number described in
Section~\ref{sec:semantic-versioning} to create a tuple that could then be used
to calculate a version delta, while another method uses the release date. The
authors only evaluated packages based on the release date and stated that there
is no meaningful way to aggregate all the version numbers together due to the
extreme variability between projects and how they assign version numbers.

To overcome the limitations of the version number delta
identified by \citet{cox_measuring_2015}, we defined
formula~\ref{eq:version_number_delta} for PVAC that uses a weighted semantic
versioning scheme to calculate the \revised{Version Number Delta (VND), a single score quantifying overall package version activity,} between two versions of a package in a
heterogeneous package management system. When combined with the Version
Categorizer as described in~\ref{sec:version_categorizer}, \revised{we derive a single numeric score enabling software maintainers and release managers to track ecosystem activity and prioritize critical dependency updates.}

\begin{equation}
    \begin{aligned}
    \label{eq:version_number_delta}
    \Delta = \sum_{i=1}^{n} ((major2_i * Q) + (minor2_i * R) + (patch2_i * S))\\ - ((major1_i * Q) + (minor1_i * R) + (patch1_i * S))
    \end{aligned}
\end{equation}

In semantic versioning, the major version signals changes that break API backward-compatibility,
the minor version signifies new functionality in a backward-compatible
way, and the patch version typically addresses backward-compatible bug
fixes~\citep{preston-werner_semantic_nodate}. Intuitively, developers often
treat major version increments as a costlier, riskier change requiring
more integration resources. Meanwhile, minor and patch increments, though less
risky, are vital because they can introduce feature
improvements, performance changes, or fix newly discovered security flaws.

\revised{
To reflect this hierarchy, VND weights these components differently, where $Q$,
$R$, and $S$ quantify the relative ``cost'' or ``impact'' of increments in the
major, minor, and patch fields, respectively. Although the semantic versioning
specification conceptually assigns a higher importance to the major field, the
minor and patch fields can still carry significant meaning for evolving software
ecosystems. For instance, a fast-paced project might release new features (minor
increments) every few weeks, whereas a “stable” project might rarely update its
major version but produce numerous critical security patches. Consequently,
weighting must balance the theoretical hierarchy of version fields against
practical observations of how the ecosystem evolves.
}

\revised{We conducted a sensitivity analysis
using a Total-Order Index (STST) to determine empirically sound weights.} In particular, we randomly sampled triplets
($Q$, $R$, $S$) and measured how changes in each weight influenced the final VND
score relative to established methods, such as using the release date.
Our analysis (detailed in Section~\ref{sec:rq2}) found that
major, minor, and patch changes each contributed a non-trivial impact to overall
software activity. However, major increments remained slightly more significant in
the optimal weighting.

\subsection{Activity Categorizer (AC)}
\label{sec:activity_categorizer}

The AC component \revised{identifies} individual packages that have not
seen any updates between releases, \revised{allowing practitioners to quickly identify specific packages that require immediate updates or vulnerability assessments.} While the
VND component gives a global view of ecosystem activity, the AC provides the
ability to drill down and see activity at the package level.
The AC component takes the version strings for the same
package across two different releases to determine the \revised{package's activity level}.

\begin{lstlisting}[
    language={Python},
    caption={Pseudocode implementation of the Activity Categorizer.},
    captionpos=b,
    label={lst:pseudo_pvac}]
def pvac(pkg_R1,pkg_R2):

    sem_R1 = lookup_sem_class(pkg_R1)
    sem_R2 = lookup_sem_class(pkg_R2)

    if sem_R1['epoch'] != sem_R2['epoch']:
        return None

    activity_level = "Sedentary"

    if sem_R1['major'] != sem_R2['major']:
        activity_level = "Very Active"
    elif sem_R1['minor'] != sem_R2['minor']:
        activity_level = "Moderately Active"
    elif sem_R1['patch'] != sem_R2['patch']:
        activity_level = "Lightly Active"

    return activity_level
\end{lstlisting}

\begin{table}[ht]
    \caption{Activity level categories and their respective descriptions.}
    \centering
    \begin{tabular}{p{0.25\linewidth} p{0.5\linewidth}}
        \toprule
         Activity Level & Description \\
        \toprule
         Very Active & Differences found in major version\\
         Moderately Active & Matching major version, but a difference was found in the minor version \\
         Lightly Active & Matching major and minor versions, but a difference was found in the patch version \\
         Sedentary & Matching major, minor, and patch versions \\
        \bottomrule
    \end{tabular}

    \label{tab:pvac_activity_classificaiton}
\end{table}

Based on the Debian versioning policy, it is only valid to compare versions
with the same epoch. If the upstream major version \revised{differs} between the two
versions, the AC component categorizes the package as \textbf{Very Active} over the period between the two releases. If the upstream major version is the same, but
the minor version has changed, AC categorizes the package as
\textbf{Moderately Active} over the evaluation period. If the upstream major and
minor versions are identical, but the patch level has changed, AC categorizes
the package as \textbf{Lightly Active}.  \revised{Finally, if both releases' upstream major,
minor, and patch versions are identical,} the package is
categorized as \textbf{Sedentary}. Table~\ref{tab:pvac_activity_classificaiton}
provides a more concise description of the AC  Categories, and  Listing
\ref{lst:pseudo_pvac} shows a pseudocode implementation of the AC component.

\subsection{Data Set Construction}
\label{sec:dataset}

\begin{figure}[htbp]
    \centering
    \includegraphics[scale=.7]{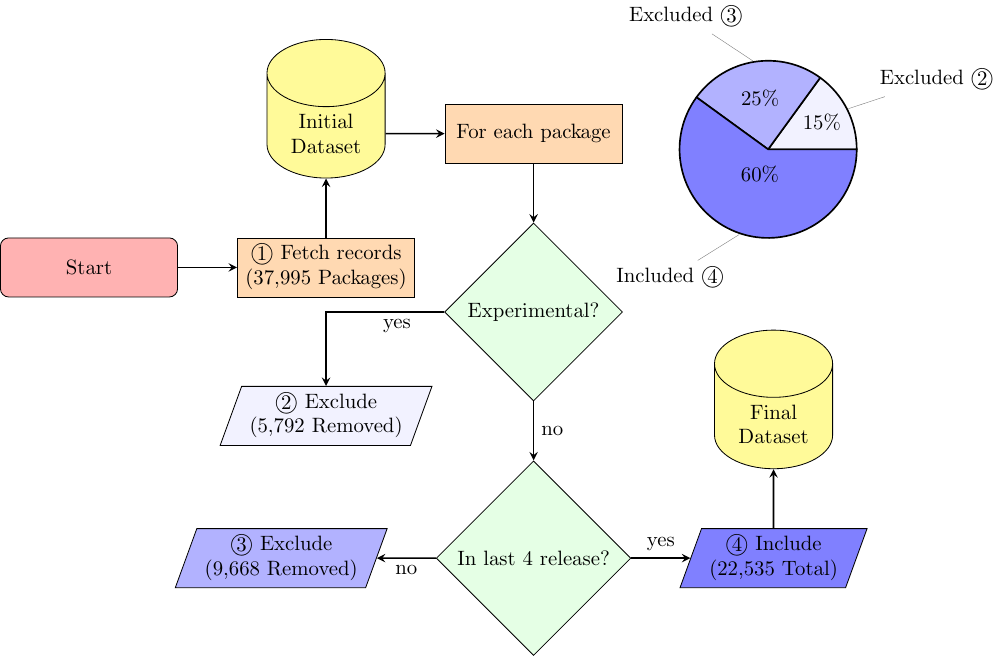}
    \caption{Methodology for constructing the dataset using the Ultimate Debian
    Database (UDD). The annotated figure shows the process of extracting the necessary
    information. \revised{The pie graph in the upper right visualizes the number of excluded and included packages from the original dataset.  Section \ref{sec:dataset_overview} describes items
    \textcircled{\raisebox{-0.8pt}{1}}, \textcircled{\raisebox{-0.8pt}{2}}, \textcircled{\raisebox{-0.8pt}{3}}, and \textcircled{\raisebox{-0.8pt}{4}} in detail.}}
    \label{fig:dataset-construction}
\end{figure}

\begin{table}[ht]
    \centering
    \caption{Distribution of the number of packages across all releases of the
    apt package manager system. The table shows the number of packages for each
    release, the total number of packages, and the mapping between
    Debian and Ubuntu release names.}
    \begin{tabular}{lllc}
    \toprule
    Release & Ubuntu & Debian & Count \\
    \midrule
    10 & 18.04 LTS & buster & 5,657 \\
    11 & 20.04 LTS & bullseye & 5,748 \\
    12 & 22.04 LTS & bookworm & 6,070 \\
    13 & 24.04 LTS & trixie & 5,060 \\
    Total &  &  & 22,535 \\
    \bottomrule
    \end{tabular}

    \label{tab:data-distribution}
\end{table}

\begin{table}[ht]
    \centering
    \caption{A selected sample of the constructed dataset. The table
    shows the UDD release number, distribution (Ubuntu and Debian), package name, version, and
    date. The raw database is stored in a CSV file and is used as input to
    PVAC.}
    \begin{tabular}{{llllll}}
    \toprule
    Release & Ubuntu & Debian & Package & Version & Date \\
    \midrule
    13 & 24.04 LTS & trixie & r-cran-spp & 1.16.0-2 & 2020-09-23 \\
    11 & 20.04 LTS & bullseye & gnome-firmware & 3.36.0-1 & 2020-12-01 \\
    10 & 18.04 LTS & buster & grap & 1.45-1 & 2016-11-22 \\
    12 & 22.04 LTS & bookworm & kasumi & 2.5-11 & 2022-12-04 \\
    10 & 18.04 LTS & buster & nullmailer & 1:2.2-3 & 2018-12-24 \\
    13 & 24.04 LTS & trixie & switcheroo-control & 2.6-3 & 2024-05-29 \\
    13 & 24.04 LTS & trixie & dnstracer & 1.9-8 & 2023-12-08 \\
    13 & 24.04 LTS & trixie & libnxt & 0.5.2-1 & 2025-02-25 \\
    12 & 22.04 LTS & bookworm & devhelp & 43.0-3 & 2022-12-19 \\
    10 & 18.04 LTS & buster & loki & 2.4.7.4-8 & 2018-10-12 \\
    \bottomrule
    \end{tabular}
    \label{tab:sample-dataset-table}
\end{table}

\subsubsection{Dataset Construction Overview}\label{sec:dataset_overview}

As of 2025, distrowatch.com\footnote{https://distrowatch.com/search.php} lists
125 out of 273 distributions, basing their package management system on apt. This
gives apt-based systems a market share of 45.8\%. Therefore, for our automated
dataset construction, we chose to focus our evaluation on distributions based on
the apt package manager system, which \revised{yields} the most significant
insights for our analysis and covers many popular Linux distributions, such as
Debian and Ubuntu. \revised{Section~\ref{sec:extending-pvac} details the minor modifications
necessary to adapt PVAC to other ecosystems, such as Fedora or Arch.}

We constructed the dataset through a systematic extraction and
filtering process, leveraging the Ultimate Debian Database (UDD)\footnote{https://wiki.debian.org/UltimateDebianDatabase/}, a
comprehensive and publicly available database for Debian package information.
Fig.~\ref{fig:dataset-construction} visually illustrates the step-by-step dataset
construction process, described in detail below. The dataset
construction involved multiple explicit stages, ensuring clarity and
reproducibility.

\begin{itemize}
\item[\textcircled{\raisebox{-0.9pt}{1}}] Fetch records

We initially queried UDD to obtain all available packages, focusing primarily on
relevant fields such as the package name, version number, release information,
and upload date. We restricted our query to only return packages 
available for the AMD64 architecture, widely used in desktop and server
environments, ensuring applicability to common production scenarios. The raw
dataset initially contained 37,995 entries before filtering.

\item[\textcircled{\raisebox{-0.9pt}{2}}] \revised{\textbf{Exclude}} Experimental

We excluded any packages labeled ``experimental'' according to Debian guidelines to ensure our dataset contains only stable packages.
Experimental packages were identified through explicit tags provided by UDD.
This filtering reduced our dataset by 5,792 packages.

\item[\textcircled{\raisebox{-0.9pt}{3}}] \revised{\textbf{Exclude not}} In the Last Four Releases

We limited the dataset to packages consistently present in the last four major
\revised{Debian releases} (and corresponding Ubuntu LTS releases). An additional 9,668
packages were removed.

\revised{\item[\textcircled{\raisebox{-0.9pt}{4}}] \textbf{Include} in Final Dataset

The final dataset contained \datasetSize{} packages. 15\% of the packages excluded
were experimental, 25\% of the packages were not included in the last four releases,
leaving us with 60\% of the original dataset to work with.}

\end{itemize}

Table~\ref{tab:data-distribution} presents the distribution of the final dataset
across four consecutive releases. These specific release pairs were selected due
to their stable and widely used status, ensuring the dataset's relevance for the practical analysis of software ecosystems. Table~\ref{tab:sample-dataset-table}
provides a randomly selected subset of ten packages from the final dataset to
illustrate its structure and content. The diversity of version formats shown in
this sample underscores the heterogeneous nature of the dataset, reinforcing the
necessity for our proposed approach.

\begin{figure}[htbp]
    \centering
    \includegraphics{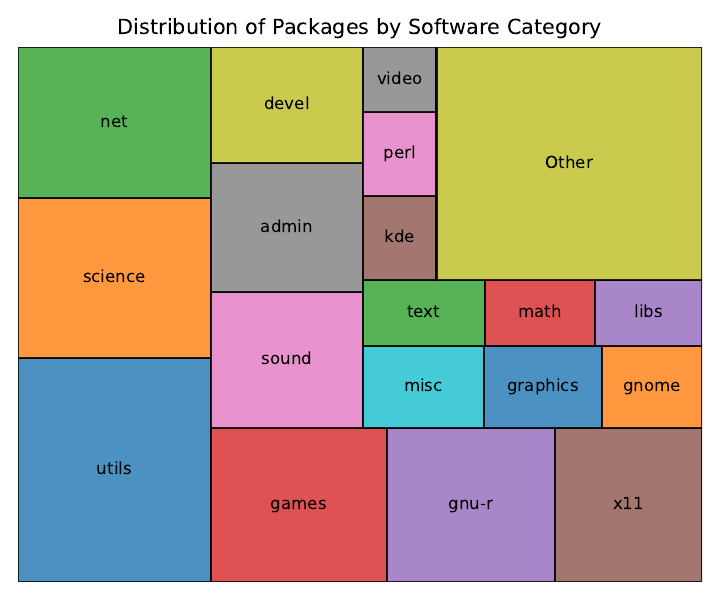}
    \caption{Distribution of the number of packages across the different software
    categories of the dataset constructed from UDD. The figure shows all categories
    with over 300 packages; the remaining categories \revised{are} grouped into the
    ``Other'' category.}
    \label{fig:section-distribution}
\end{figure}

The packages in the dataset are a diverse mix of software categories
written in various programming languages, representing the heterogeneous
nature of all the software used to construct a complete working system. The
dataset contains packages representing software from developer tools, user interfaces, and system utilities. Fig.~\ref{fig:section-distribution} visually shows the relative
distribution of packages across the different software categories.
The `utils' category has the most packages, followed by `science' and `net'.
\revised{For clarity, software} categories with \revised{fewer} than 300 packages are grouped into the
`Other' category.

\section{Results}
\label{sec:results}

Having described our methodology and dataset, we will now use both to answer the
research questions in Section~\ref{research_questions}.

\subsection{\textbf{RQ1: \RQone}}
\label{sec:rq1}

\begin{figure}[htbp]
    \centering
    \includegraphics[width=\columnwidth]{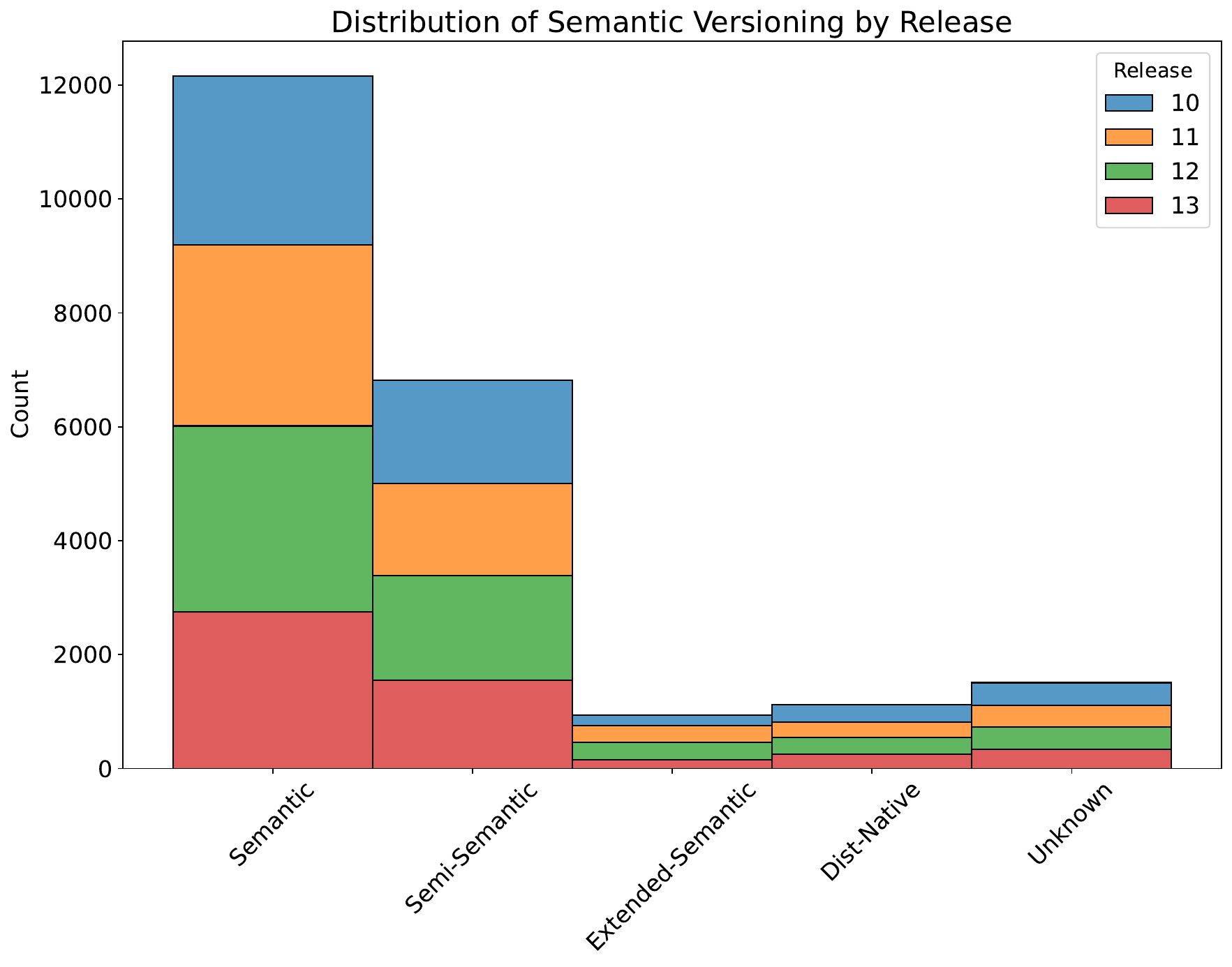}
    \caption{Distribution of categorized package versions across four semantic
    categories, highlighting the prevalence of Semantic and Semi-Semantic
    versioning in Debian-based package ecosystems. Each column shows the total
    number of packages from each release across the four categories.}
    \label{fig:rq1-results}
\end{figure}

We evaluated PVAC's ability to categorize all \datasetSize{}
packages. Table~\ref{tab:rq1-confusion-matrix} shows that PVAC had a  True
Positive (TP) rate of \tprate{}, a False Positive (FP) rate of 1,335, a True
Negative (TN) rate of 71, and a False Negative (FN) rate of 102. This
gives PVAC a precision of 93.87\% and a recall of 99.52\%. This validates the
effectiveness of the Version Categorizer component. Additionally, as shown in
Fig.~\ref{fig:rq1-results}, packages within the apt package manager dataset
overwhelmingly follow the standard semantic versioning scheme.

Table~\ref{tab:rq1-sample-dataset} shows a sample of the results of
running PVAC on the apt package manager dataset. The
table shows the package, version, categories, and extracted numbers from the
semantic version.

\begin{table}[htbp]
    \centering\
    \caption{Randomly selected sample of results after running the Version
    Categorizer on the apt package manager dataset. The table shows the package,
    version, categories, and extracted numbers from the semantic version.}
    \centering
    \resizebox{\columnwidth}{!}{%
    \begin{tabular}{{p{.75cm}lllllllp{0.3cm}p{0.3cm}p{0.3cm}}}
    \toprule
    Release & Package & Version & Semantic & Major & Minor & Patch \\
    \midrule
    11 & bedtools & 2.30.0+dfsg-1 & Dist-Native & 2 & 30 & 0 \\
    11 & authbind & 2.1.2 & Semantic & 2 & 1 & 2 \\
    13 & rebar & 2.6.4-4 & Semantic & 2 & 6 & 4 \\
    10 & cervisia & 4:17.08.3-1 & Semi-Semantic & 17 & 8 & 3 \\
    11 & semodule-utils & 3.1-1 & Semi-Semantic & 3 & 1 & 0 \\
    12 & libdata-peek-perl & 0.52-1 & Semi-Semantic & 0 & 52 & 0 \\
    11 & r-cran-relsurv & 2.2-3-2 & Semi-Semantic & 2 & 2 & 0 \\
    12 & dhis-client & 5.5-6 & Semi-Semantic & 5 & 5 & 0 \\
    13 & ksmbd-tools & 3.5.2-3 & Semantic & 3 & 5 & 2 \\
    12 & pilercr & 1.06+dfsg-5 & Dist-Native & 1 & 6 & 0 \\
    12 & wob & 0.14.2-1 & Semantic & 0 & 14 & 2 \\
    11 & hexalate & 1.1.6-1 & Semantic & 1 & 1 & 6 \\
    12 & kexi & 1:3.2.0-3 &Extended-Semantic & 3 & 2& 0& \\
    12 & agedu & 20211129.8cd63c5-1 & Unknown & 0 &0 &0\\
    \bottomrule
    \end{tabular}
    }
    \label{tab:rq1-sample-dataset}
\end{table}

\begin{table}[htbp]
    \centering\
    \caption{Confusion matrix of the Version Categorizer results after running on
    our dataset. The table shows the True Positive (TP), False Positive (FP), True
    Negative (TN), and False Negative (FN) rates.}
    \centering
    \begin{tabular}{lcc}
        \toprule
        & \textbf{Predicted Positive (PP)} & \textbf{Predicted Negative (PN)} \\
        \midrule
        \textbf{Positive (P)} & 21,027 (TP) & 102 (FN)\\
        \textbf{Negative (N)} & 1,335 (FP) & 71 (TN)\\
        \bottomrule
    \end{tabular}
    \label{tab:rq1-confusion-matrix}
\end{table}

\begin{tcolorbox}[
    enhanced,attach boxed title to top center={yshift=-3mm,yshifttext=-1mm},
    colback=blue!5!white,colframe=blue!75!black,colbacktitle=red!80!black,
    title=Answer to RQ1,fonttitle=\bfseries,
    boxed title style={size=small,colframe=red!50!black} ]

    For systems based on the apt package manager, we found that PVAC
    successfully categorized \tprate{} out of \datasetSize{} packages
    in our dataset. The distribution of the packages across the four categories
    is shown in Fig.~\ref{fig:rq1-results}. Additionally, we found
    that most of the packages in the dataset follow the Semantic and
    Semi-Semantic versioning scheme. Thus, semantic versioning is the primary
    versioning strategy observed in our dataset.
\end{tcolorbox}

\subsection{\textbf{RQ2: \RQtwo}}
\label{sec:rq2}

First, we  determined the weights $Q$, $R$, and $S$ for
formula~\ref{eq:version_number_delta} by running a sensitivity analysis using a
Total-Order Index (ST) with Table~\ref{tab:rq2-results-table} showing the
optimal weights that were derived. We found that each component
of the semantic version number showed equal significance, with the major version number only slightly higher than the minor and patch. While the major
version is typically the most important component due to the breaking changes
that are introduced, the minor and patch versions are also shown to be highly
important, as they can introduce new features and bug fixes.

\begin{figure}[htbp]
    \centering
    \includegraphics[width=\columnwidth]{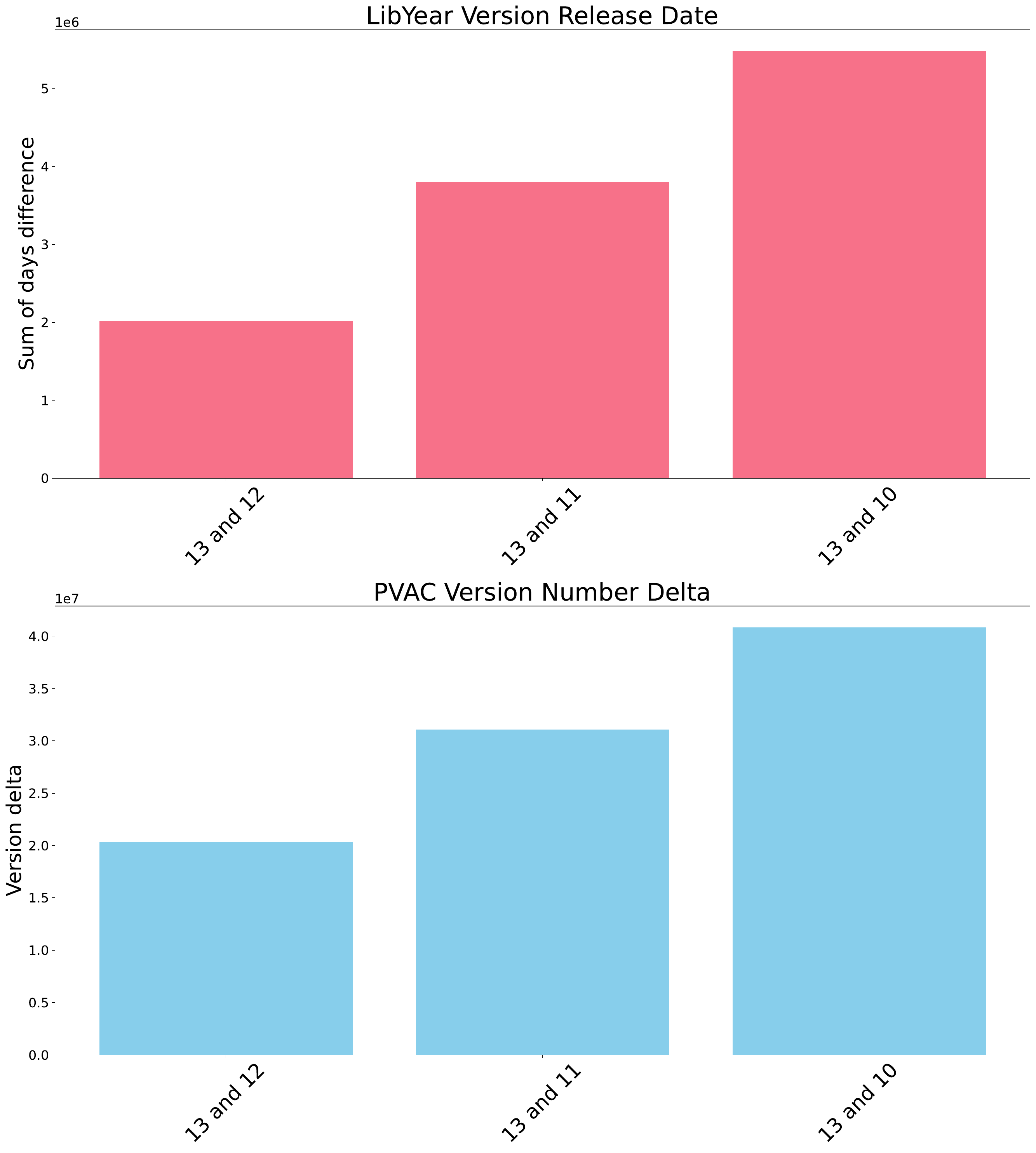}
    \caption{Results of evaluating the VND Score compared to the Libyears metric
    using the release date of the package. The graph shows both the VND Score
    and the Libyears metric as we go back in time, and shows that the VND
    component is consistent with the Libyears metric.}
    \label{fig:rq2-results}
\end{figure}

To validate the VND component and weights, we compared the VND Score to the
Libyears metric described by \citet{cox_measuring_2015}.
Fig.~\ref{fig:rq2-results} shows that as we compare the latest release
with older and older releases, we correctly detect activity in the system as
packages get farther and farther out of date. This shows that the VND component
is consistent with an existing metric and can detect activity within the
system. Thus, we have validated our VND metric and released it to the
community to measure the activity of a heterogeneous package management
system using only semantic version information.

\begin{table}[htbp]
    \caption{Sensitivity analysis results of the VND component using
    the Total-Order Index (ST) and Total-Order Index Confidence. The table shows
    the resulting weights for the major ($Q$), minor ($R$), and patch ($S$)
    that were defined in equation~\ref{eq:version_number_delta}.}
    \centering
    \begin{tabular}{{lcc}}
    \toprule
    Parameter & Total-Order Index (ST) & Total-Order Index Confidence (ST\_conf) \\
    \midrule
    major & 0.3373 & 0.0212 \\
    minor & 0.3306 & 0.0201 \\
    patch & 0.3321 & 0.0223 \\
    \bottomrule
    \end{tabular}
    \label{tab:rq2-results-table}
\end{table}

\begin{tcolorbox}[
    enhanced,attach boxed title to top center={yshift=-3mm,yshifttext=-1mm},
    colback=blue!5!white,colframe=blue!75!black,colbacktitle=red!80!black,
    title=Answer to RQ2,fonttitle=\bfseries,
    boxed title style={size=small,colframe=red!50!black} ]

   The VND component of PVAC successfully calculated a single score
    representing the activity of a heterogeneous package management system. We
    show that PVAC performs exceptionally well at surfacing ecosystem activity within our dataset
    compared to existing metrics. Additionally, PVAC can automatically
    determine the optimal weights for the major, minor, and patch components,
    allowing it to adapt to future changes in the system over time.

\end{tcolorbox}

\subsection{\textbf{RQ3: \RQthree}}
\label{sec:rq3}

\begin{figure}[htbp]
    \centering
    \includegraphics[width=1.0\linewidth]{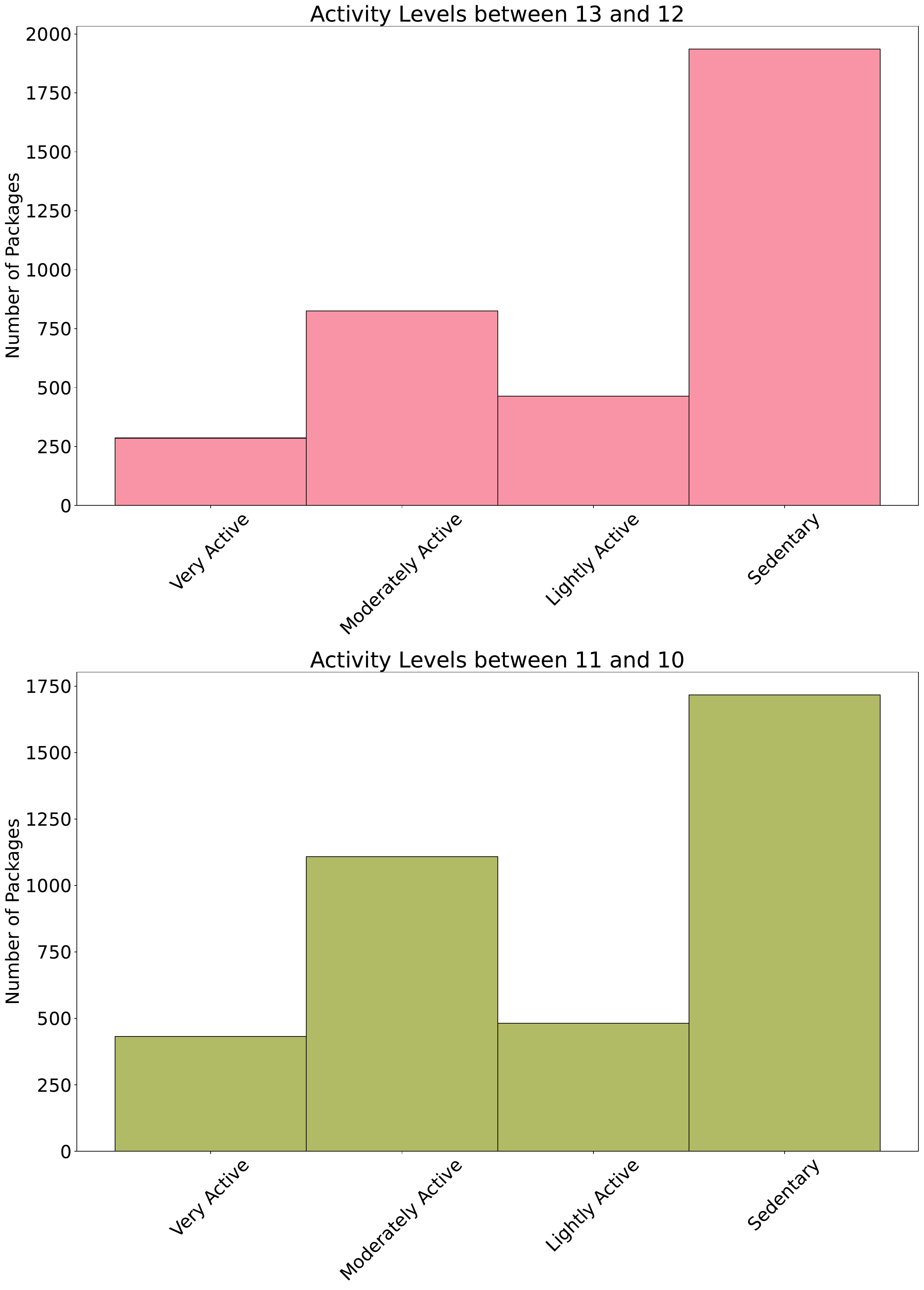}
    \caption{Results of running the Activity Categorizer against each
    sequential pairwise release of the apt dataset. The histogram shows the
    distribution of the packages across the four categories and shows that the
    majority of the packages are categorized as Sedentary between releases.}
    \label{fig:rq3-results}
\end{figure}

\revised{To} evaluate the effectiveness of the Activity Categorizer component, we
did a pairwise comparison of packages between major releases in the dataset. It
should go without saying that any new package that was introduced would be
excluded from this analysis, as it would not have a previous release against
which to compare. Fig.~\ref{fig:rq3-results} shows the distribution of
the packages across the four categories, with the numbers shown in
Table~\ref{tab:rq3-results-table}. Table~\ref{tab:rq3-sample-table}
shows a sample of the results of the pairwise comparison between release 12
and release 13 of the apt dataset.

\begin{table}[htbp]
    \caption{Results of running the Activity Categorizer on the apt package
    manager dataset, comparing the activity between releases. The table shows only
    packages that were common between the release pairs and were successfully
    categorized.}
    \centering
    \begin{tabular}{l p{1cm} p{1.5cm} p{1cm} p{1.5cm} p{1cm} p{1cm}}
    \toprule
    Activity & Very Active & Moderately Active & Lightly Active & Sedentary & Total \\
    \midrule
    13 and 12 & 281 & 807 & 452 & 1,911 & 3,451 \\
    11 and 10 & 419 & 1,092 & 461 & 1,692 & 3,664 \\
    \bottomrule
    \end{tabular}

    \label{tab:rq3-results-table}
\end{table}

\begin{table}[htbp]
    \centering
    \caption{A selected sample of results after running the Activity
    Categorizer on the apt package manager dataset showing a pairwise comparison
    of releases \revised{and their respective domains}.}
    \resizebox{\columnwidth}{!}{%
    \begin{tabular}{lllll}
    \toprule
    Package & Version 13 & Version 12 & Activity & \revised{Domain} \\
    \midrule
    lomiri-history-service & 0.6-2 & 0.4-2 & Moderately Active & \revised{libs} \\
    pipewalker & 1.1-1 & 0.9.5-2 & Very Active & \revised{games}  \\
    bruteforce-luks & 1.4.0-4 & 1.4.0-4 & Sedentary & \revised{utils} \\
    musl & 1.2.5-1.1 & 1.2.3-1 & Lightly Active & \revised{libs}  \\
    xinput & 1.6.4-1 & 1.6.3-1 & Lightly Active & \revised{x11} \\
    mapsembler2 & 2.2.4+dfsg1-4 & 2.2.4+dfsg1-4 & Sedentary & \revised{science} \\
    dos2unix & 7.5.2-1 & 7.4.3-1 & Moderately Active & \revised{text} \\
    r-cran-tfmpvalue & 0.0.9-1 & 0.0.9-1 & Sedentary & \revised{gnu-r} \\
    tk8.6 & 8.6.15-1 & 8.6.13-2 & Lightly Active & \revised{interpreters} \\
    simrisc & 16.02.00-1 & 15.00.00-1 & Very Active & \revised{science} \\
    gnome-flashback	& 3.54.0-1	& 3.46.0-1	& Moderately Active & \revised{gnome} \\
    ghc	& 9.6.6-3	& 9.0.2-4	& Moderately Active & \revised{haskell}\\
    gnome-music	& 47.0-1 & 42.1-1 & Very Active & \revised{gnome}\\
    squeekboard	& 1.42.1-2	& 1.21.0-1	& Moderately Active& \revised{x11}\\
    openjfx	& 11.0.11+1-3.2	& 11.0.11+1-3 & Sedentary& \revised{java}\\
    
    \bottomrule
    \end{tabular}
    }
    \label{tab:rq3-sample-table}
\end{table}

\revised{\subsection{Distinguishing Stability and Staleness}\label{sec:rq3-stability-staleness}}

While the Activity Categorizer can categorize the activity of individual packages, 
it can not provide reasons
for the lack of activity. For example, a package management system
may prioritize stability and only update packages when necessary.
Our research into the apt package manager system shows that the majority of the
packages are categorized as Sedentary between releases, which is consistent with
Debian's policy of prioritizing
stability~\citep{hertzogDebianAdministratorsHandbook2012} over bleeding-edge
releases.

In practice, a Sedentary classification can arise for at least two
distinct reasons:

\begin{itemize}
    \item Stable Packages - Some packages are intentionally
    stable or feature-complete. For
    instance, in our dataset, the package \texttt{bruteforce-luks}
    (Table~\ref{tab:rq3-sample-table}) remained at the same version between
    releases 12 and 13 with no incremental changes. A closer inspection of its
    GitHub repository\footnote{https://github.com/glv2/bruteforce-luks} shows
    bug fixes as recently as 2024 and active interaction with user-reported
    issues and bugs. Thus, it is likely feature-complete and is still actively
    maintained. Such packages typically still see sporadic updates (e.g.,
    security patches), but the release cadence remains low.

    \item Stale or Abandoned Packages - In other cases, remaining
    on the same version number may indicate a lack of active maintenance. For
    instance, \texttt{mapsembler2} (Table~\ref{tab:rq3-sample-table}) was also
    unchanged between releases 12 and 13. However, investigating its public
    homepage\footnote{https://colibread.inria.fr/software/mapsembler2/} revealed
    that it is no longer maintained, with the last update in 2014. An orphaned or
    stale package may accrue unaddressed technical debt, fail to fix
    security vulnerabilities, or no longer respond to community needs.

\end{itemize}

In a production context, such as continuous integration pipelines,
projects using Sedentary dependencies might impose additional checks: for
instance, regularly scraping repository details or issue-tracker data to see if a
package marked Sedentary receives the minimal updates needed to
remain safe. If not, engineers can proactively migrate away from that
dependency or allocate resources to maintain it.

\begin{tcolorbox}[
    enhanced,attach boxed title to top center={yshift=-3mm,yshifttext=-1mm},
    colback=blue!5!white,colframe=blue!75!black,colbacktitle=red!80!black,
    title=Answer to RQ3,fonttitle=\bfseries,
    boxed title style={size=small,colframe=red!50!black} ]

We evaluated the effectiveness of the Activity Categorizer component and found that it could categorize the activity of
individual packages in a heterogeneous
package management system by comparing packages that are common between
release pairs. Our research shows that most packages were categorized as Sedentary
in the apt package manager dataset, which \revised{is consistent} with existing research~\citep{legayPackageFreshnessLinux2020}.
\end{tcolorbox}
\section{Discussion}
\label{sec:discussion}

Software is complex and is built on a foundation of many software
packages, many of which may not be directly linked to the final application being
developed. It is \revised{essential} to look at every piece of software required
to make a system function, even if a direct link to the final application is not
apparent. In this paper, we have presented a way to tackle this problem
by categorizing the semantic versioning information and then using that
information to derive activity at both the ecosystem and package levels.

\subsection{Research Questions}

Each research question surfaced interesting findings. In \textbf{RQ1}
(Section~\ref{sec:rq1}), we found that the majority of packages in our dataset
follow semantic versioning. Additionally, we found that with only minor
modifications to the official semantic standard, our tool could categorize
\tprate{} out of \datasetSize{} packages in our dataset. Our findings closely
align with~\citep{dietrichDependencyVersioningWild2019}, highlighting that
heterogeneous software ecosystems widely adopt semantic and flexible versioning
practices. Our research extends these insights by introducing finer-grained
categorization beyond the traditional semantic versioning scheme.

In \textbf{RQ2} (Section~\ref{sec:rq2}), we found that the VND component
is a useful metric for assessing the activity of the entire ecosystem. We
found that the Version Number Delta can be used to generate a single number that
can be used to keep a pulse on the activity of an ecosystem, and we validated
our findings against established metrics.

For \textbf{RQ3} (Section~\ref{sec:rq3}), we used the AC component to label the activity of
individual packages in our dataset. We found that most packages derived from the Debian package manager are Sedentary, meaning the version
number does not change often. This finding provides empirical evidence to
support existing research~\citep{nguyenLifeDeathSoftware2012,
legay_quantitative_2021, legayPackageFreshnessLinux2020}. As noted in
Section~\ref{sec:rq3-stability-staleness}, the Sedentary label is not
necessarily a bad thing as it could mean a package is feature-complete but still
maintained as is the case with the \texttt{bruteforce-luks} package, or it could
indicate a package has been abandoned and will never see any future updates as
evidenced by the \texttt{mapsembler2} package (Table~\ref{tab:rq3-sample-table}). PVAC can not distinguish between these two
cases; it is up to the user to determine the current state of the package.

It is important to note that PVAC strongly depends on time. Our research was
conducted using discrete releases from the apt package manager. This means a package labeled Sedentary may have been updated since our last data
collection. This is a limitation of our \revised{research, not} of the
tool itself.  For example, if PVAC were to be used in a CI/CD pipeline, the data
would be collected in real-time, and the results would be more accurate. This
would help distinguish between packages that are truly sedentary and packages
that are just between updates.

The Activity Categorizer provides empirical evidence to consumers of a package
management system that packages are either actively updated or not. Our large-scale study confirms other researchers' results regarding the stability of apt-based systems
\citep{legayPackageFreshnessLinux2020,legay_quantitative_2021}. If a company wants to adopt a package management system that prioritizes stability over the
latest features, then the Activity Categorizer can provide the necessary
information to make that decision. \revised{Additionally, we found that the
semantic versioning standard's major, minor, and patch portions} encode
equally important information about the entire system, with the major
version being only slightly more important than the minor and patch versions.

\revised{
\subsection{Extending PVAC}\label{sec:extending-pvac}
}

Although we focused on apt-based systems, PVAC is extensible to other Linux
distributions or package managers (e.g., Fedora, Arch, or others). These
ecosystems differ primarily in how they label or store version metadata, but
PVAC's core logic remains the same:

\begin{description}
    \item[\textbf{Data Collection}]- Fedora and other RPM-based distributions store
    metadata in repositories that can be queried through commands like
    \texttt{dnf} or \texttt{Koji}. Arch Linux similarly provides tar archives in
    its package repositories. By extracting each package's name, version,
    release date (or build date), and architecture, we could produce a CSV with
    the minimal fields required by PVAC.

    \item[\textbf{Version Categorization}]- The regular expressions from
    Section~\ref{sec:version_categorizer} can be adapted to each packaging
    system's version conventions. For instance, Fedora's RPM versions can
    contain an Epoch (similar to Debian's) and a Release field that includes
    distribution-specific increments. We would create an Extended-Semantic or
    Dist-Native variant to handle those distribution-tailored fields.

    \item[\textbf{Version Number Delta}]- Once we parse out the major, minor, patch (and
    optional epoch) from each version label, the VND formula applies exactly as
    in apt-based systems.

    \item[\textbf{Activity Categorizer}]-  As long as the same package can be identified
    across multiple releases, we can label it using Very Active, Moderately
    Active, Lightly Active, or Sedentary in precisely the same manner.
\end{description}

\revised{
In short, adapting PVAC to Fedora, Arch, or other ecosystems requires only a
small amount of custom code to retrieve and parse version strings from that
ecosystem's repositories. The underlying categorization logic and VND weighting
remain unchanged.}

\revised{\subsection{Comparison with Alternative Approaches}\label{sec:alternative-approaches}}

A natural question is whether a rule-based system such as ours might be
replaced or substantially improved by a machine learning (ML) classifier. Below,
we briefly discuss how such methods compare and why we selected a hybrid of
regular expressions and weight-based scoring for version activity analysis:

\begin{description}
    \item[\textbf{ML for Version String Parsing}] - An ML model could
    theoretically learn to distinguish semantic version strings from ad hoc
    formats. However, the multi-token format of version strings often follows
    well-defined patterns (e.g., Major.Minor.Patch, possibly with appended
    tags). This makes standard regular expressions surprisingly
    robust at near-zero cost; an ML model would need a manually labeled dataset
    and would likely be overkill to parse a pattern that is already heavily
    standardized.

    \item[\textbf{ML for Package Activity}] - Another perspective is to predict whether a
    package is ``active'' by training on features like upstream commit velocity,
    distribution release cycles, or developer community signals (e.g., number of
    open PRs). However, these approaches require extensive metadata beyond the
    raw version strings (e.g., repository commits, contributor profiles), which
    are not always standardized or easy to extract across different
    distributions and would require ongoing maintenance. Our objective here is
    narrower: track version number changes as an easily measurable proxy of
    activity.

    \item[\textbf{Justification for a Rule-Based and Weighted Approach}] - In many
    large-scale OS package ecosystems, version strings are explicitly documented
    (e.g., Debian Policy Manual, Fedora Packaging Guidelines). A rule-based
    approach capitalizes on this inherent structure, offering interpretable
    results. By contrast, an ML classifier might yield high accuracy for
    known distributions but lack interpretability, hamper user trust, and 
    require frequent retraining as new versioning schemes arise.

\end{description}

\revised{
Thus, while machine learning classifiers can provide complementary insights, we
conclude that a lightweight, distribution-centric, and rule-based system is the
most direct, transparent approach for measuring version activity across
heterogeneous package managers. Future work could explore blending additional
signals such as commits, issue trackers, or community metrics for a more nuanced
assessment.}

\subsection{Practical Implications and Recommendations}

When building a complete project for release or deployment, a company could use
our methods and associated tools in a continuous integration and deployment
(CI/CD) pipeline such as GitHub
Actions\footnote{https://github.com/features/actions}. \revised{Software release managers can define acceptable Version Number Delta thresholds to determine the necessary level of acceptance testing before deployment.} If the Version Number Delta is high, there
is a high likelihood that the dependencies have been updated \revised{significantly}, and the project
should \revised{undergo thorough testing. Conversely, if the Version Number Delta is low, 
the dependencies have changed minimally since the last release, indicating the
project can be safely deployed with reduced testing. Integrating PVAC into CI/CD pipelines allows teams to streamline testing processes, minimizing the resources required for deployment, which is especially valuable in modern DevOps environments where software deployments happen frequently and cloud resources incur per-minute charges.}

An example set of recommendations for practitioners and researchers \revised{is as follows:}

\begin{itemize}
    \item Automatically run PVAC when dependencies are updated or at scheduled
    intervals.
    \item Set alert thresholds for VND scores to trigger manual review if
    the score exceeds acceptable limits.
    \item Periodically use Activity Categorizer outputs to \revised{assess and proactively} replace Sedentary packages that may pose security risks due to
    inactivity.
\end{itemize}

Finally, we recommend further integrating PVAC's outputs with existing security and
maintenance dashboards to streamline developer decisions on dependency
management.
\section{Threats to Validity}
\label{sec:threats}

In this section, we discuss threats to validity relevant to our study,
categorized into external and internal threats.

\subsection{External Validity}
\label{sec:internal}

We constructed our dataset from the Ultimate Debian Database (UDD), relying on
its data accuracy and comprehensiveness. Despite the database's stated goal of
``faithful gathering,'' previous research has identified potential omissions or
inaccuracies in UDD
data~\citep{nussbaumUltimateDebianDatabase2010, daviesPerspectivesBugsDebian2010}.
Additionally, although we systematically filtered experimental and unstable
packages to ensure our dataset represented stable releases, it is possible that some
misclassifications remained, potentially influencing the generalizability of our
findings. \revised{It is important to note that due to our reliance on UDD, PVAC has not been validated on other distributions. We detailed the steps to extend PVAC to other distributions in Section~\ref{sec:extending-pvac}; more work is required to validate PVAC as a general-purpose solution.}

When randomly sampling our dataset for manual verification, we found changes in the upstream versioning format between releases, and there was
neither documentation nor justification for the change. The epoch
value was not set, leading to incorrect version comparisons. This impacted the
Version Number Delta analysis in Section~\ref{sec:rq2} \revised{because} the differences
between the varying versioning schemes were sometimes invalid.

When writing our script to \revised{construct the dataset automatically}, we found a
wide variety of ways that projects versioned their software. Some projects that
we examined did have a documented method for incrementing the version
number, while other projects versioned the numbers based on the ``feeling'' of
the lead developer, while still others had no documented process on how their
packages were versioned at all.

\subsection{Internal Validity}

Our method relies heavily on regular expressions to categorize package
versioning schemes. Although we carefully constructed and manually validated
these regular expressions on random samples of our dataset, there remains a risk
that version strings with uncommon or undocumented formats could have been incorrectly categorized or missed entirely. Future improvements could include
crowd-sourced or community-based validation to increase coverage and accuracy
further. Such misclassifications could lead to inaccuracies in evaluating
version activity, impacting the validity of our results. Additionally, we
carefully annotated our regular expressions on a small sample size to
calculate the TP, FP, TN, and FN rates; it is possible that our sample size was
not large enough to capture all the possible cases.

PVAC  may occasionally  suffer from incorrect categorization of packages in the Dist-Native
category due to overlaps with other categories, such as Semantic, Extended-Semantic, 
and Semi-Semantic. Because packages
labeled as Dist-Native are specific to a distribution and lack a clear upstream source, their versioning can sometimes mimic or closely resemble other established semantic
patterns. This resemblance can lead to potential misclassifications, reducing the
accuracy of distinguishing truly native distribution packages from those employing
adapted semantic conventions. Consequently, careful consideration and possibly
additional refinement of the classification criteria may be necessary to mitigate
these overlaps and ensure accurate categorization.

The empirical approach to determining weights for the VND component is
based on a sensitivity analysis using a Total-Order Index (ST). While our
analysis provided robust empirical support for selecting weights, variations in
software development practices or ecosystem dynamics might necessitate
periodic recalibration. Continuous monitoring and adaptation of these weights
could enhance the accuracy and reliability of the VND formula.

The AC component relies on pairwise comparisons between releases. It
cannot distinguish between packages that are feature-complete but still
maintained, and those that are genuinely abandoned or stale without additional
analysis. Thus, the AC component should be supplemented by qualitative
assessments to make informed decisions regarding package maintenance and
stability.

Lastly, mapping releases between Debian and Ubuntu was based on publicly
available documentation and release schedules. \revised{Ubuntu is synchronized\footnote{https://ubuntu.com/about/release-cycle} to Debian every 6 months, with 80\% of the packages being brought over unmodified from Debian\footnote{https://wiki.ubuntu.com/Ubuntu/ForDebianDevelopers}. In some cases, packages in Ubuntu may be modified to fix policy differences between the two projects, fix bugs, or introduce newer versions of packages. These modifications could introduce potential inaccuracies in our mapping, affecting the accuracy of our analysis.}

\section{Conclusion and Future Work}
\label{sec:conclusion}

In conclusion, our research underscores the significance of monitoring the
activity of heterogeneous package manager systems. \revised{Leveraging PVAC provides
valuable insights into the open-source communities' maturity, resilience, and responsiveness.} Our methods and tools can significantly enhance the
understanding and monitoring of ecosystem activity, providing actionable
insights for researchers and practitioners aiming to maintain secure, reliable,
and up-to-date software systems. \revised{Future work should validate PVAC
across additional ecosystems and extend its functionality to predict potential
dependency abandonment without manual intervention.}

As a result of this research, we have identified several areas for future work.
\revised{Our finding that the semantic versioning's major, minor, and patch portions 
all hold relatively equal importance in the system as a whole
suggests that further research could be done to support this finding.}
Additionally, given that semantic versioning is relatively new, it would be
interesting to see if the adoption of semantic versioning has improved
the problem of ``Dependency Hell'' as it was \revised{initially} intended, or just
added another layer of complexity to an already complex process. Finally, we
believe that the construction of an automated tool to increment the semantic
versioning of a package would be a valuable contribution to the community, as it
currently requires human intervention to determine the \revised{next} version number.

\section*{Data Availability Statement}
\label{sec:data}

We made our dataset available at:

\href{https://doi.org/10.6084/m9.figshare.28691561}{https://doi.org/10.6084/m9.figshare.28691561}

\section*{AI Usage}

The authors declare that this paper was not written by any AI or machine
learning models.



\section*{Compliance with Ethical Standards}
\textit{Conflict of Interest:} None \\

\noindent
\textit{Funding:} None \\

\noindent
\textit{Ethical approval:} Not Applicable \\

\noindent
\textit{Informed consent:} Not Applicable \\

\noindent
\textit{Clinical trial number:} Not Applicable \\

\noindent
\textit{Author Contributions:}
All authors contributed to the conception and design of the study.
All authors contributed to the manuscript. 
All authors read and approved the final manuscript.

\bibliographystyle{spbasic}
\bibliography{paper}

\begin{thebibliography}{35}
\providecommand{\natexlab}[1]{#1}
\providecommand{\url}[1]{{#1}}
\providecommand{\urlprefix}{URL }
\expandafter\ifx\csname urlstyle\endcsname\relax
  \providecommand{\doi}[1]{DOI~\discretionary{}{}{}#1}\else
  \providecommand{\doi}{DOI~\discretionary{}{}{}\begingroup
  \urlstyle{rm}\Url}\fi
\providecommand{\eprint}[2][]{\url{#2}}

\bibitem[{Avelino et~al.(2019)Avelino, Constantinou, Valente, and
  Serebrenik}]{avelinoAbandonmentSurvivalOpen2019}
Avelino G, Constantinou E, Valente MT, Serebrenik A (2019) On the abandonment
  and survival of open source projects: {{An}} empirical investigation. In:
  2019 {{ACM}}/{{IEEE International Symposium}} on {{Empirical Software
  Engineering}} and {{Measurement}} ({{ESEM}}), IEEE, Porto de Galinhas,
  Recife, Brazil, pp 1--12, \doi{10.1109/ESEM.2019.8870181}

\bibitem[{Cox et~al.(2015)Cox, Bouwers, Eekelen, and
  Visser}]{cox_measuring_2015}
Cox J, Bouwers E, Eekelen MV, Visser J (2015) Measuring {{Dependency
  Freshness}} in {{Software Systems}}. In: 2015 {{IEEE}}/{{ACM}} 37th {{IEEE
  International Conference}} on {{Software Engineering}}, IEEE, Florence,
  Italy, pp 109--118, \doi{10.1109/ICSE.2015.140}

\bibitem[{Dann et~al.(2023)Dann, Hermann, and
  Bodden}]{dannUPCYSafelyUpdating2023}
Dann A, Hermann B, Bodden E (2023) {{UPCY}}: {{Safely Updating Outdated
  Dependencies}}. In: 2023 {{IEEE}}/{{ACM}} 45th {{International Conference}}
  on {{Software Engineering}} ({{ICSE}}), pp 233--244,
  \doi{10.1109/ICSE48619.2023.00031}

\bibitem[{Davies et~al.(2010)Davies, Zhang, Nussbaum, and
  German}]{daviesPerspectivesBugsDebian2010}
Davies J, Zhang H, Nussbaum L, German DM (2010) Perspectives on bugs in the
  {{Debian}} bug tracking system. In: 2010 7th {{IEEE Working Conference}} on
  {{Mining Software Repositories}} ({{MSR}} 2010), pp 86--89,
  \doi{10.1109/MSR.2010.5463288}

\bibitem[{Decan and Mens(2021)}]{decanWhatPackageDependencies2021}
Decan A, Mens T (2021) What {{Do Package Dependencies Tell Us About Semantic
  Versioning}}? IEEE Transactions on Software Engineering 47(6):1226--1240,
  \doi{10.1109/TSE.2019.2918315}

\bibitem[{Decan et~al.(2018)Decan, Mens, and
  Constantinou}]{decanEvolutionTechnicalLag2018}
Decan A, Mens T, Constantinou E (2018) On the evolution of technical lag in the
  npm package dependency network. In: 2018 {{IEEE}} International Conference on
  Software Maintenance and Evolution ({{ICSME}}), IEEE, pp 404--414

\bibitem[{Dietrich et~al.(2019)Dietrich, Pearce, Stringer, Tahir, and
  Blincoe}]{dietrichDependencyVersioningWild2019}
Dietrich J, Pearce D, Stringer J, Tahir A, Blincoe K (2019) Dependency
  {{Versioning}} in the {{Wild}}. In: 2019 {{IEEE}}/{{ACM}} 16th
  {{International Conference}} on {{Mining Software Repositories}} ({{MSR}}),
  pp 349--359, \doi{10.1109/MSR.2019.00061}

\bibitem[{Dijkers et~al.(2018)Dijkers, Sincic, Wasankhasit, and
  Jansen}]{dijkersExploringEffectSoftware2018}
Dijkers J, Sincic R, Wasankhasit N, Jansen S (2018) Exploring the effect of
  software ecosystem health on the financial performance of the open source
  companies. In: Proceedings of the 1st {{International Workshop}} on
  {{Software Health}}, Association for Computing Machinery, New York, NY, USA,
  {{SoHeal}} '18, pp 48--55, \doi{10.1145/3194124.3194126}

\bibitem[{GNU/Linux(2014)}]{debiangnu/linuxCVE20146271GNUBourneAgain2014}
GNU/Linux D (2014) {{CVE-2014-6271 GNU Bourne-Again Shell}} ({{Bash}})
  `{{Shellshock}}' {{Vulnerability}}.
  https://nvd.nist.gov/vuln/detail/cve-2014-6271

\bibitem[{Goggins et~al.(2021)Goggins, Lumbard, and
  Germonprez}]{goggins_open_2021}
Goggins S, Lumbard K, Germonprez M (2021) Open {{Source Community Health}}:
  {{Analytical Metrics}} and {{Their Corresponding Narratives}}. In: 2021
  {{IEEE}}/{{ACM}} 4th {{International Workshop}} on {{Software Health}} in
  {{Projects}}, {{Ecosystems}} and {{Communities}} ({{SoHeal}}), pp 25--33,
  \doi{10.1109/SoHeal52568.2021.00010}

\bibitem[{{Gonzalez-Barahona}(2020)}]{gonzalez-barahonaCharacterizingOutdatenessTechnical2020}
{Gonzalez-Barahona} JM (2020) Characterizing outdateness with technical lag: An
  exploratory study. In: Proceedings of the {{IEEE}}/{{ACM}} 42nd
  {{International Conference}} on {{Software Engineering Workshops}},
  Association for Computing Machinery, New York, NY, USA, {{ICSEW}}'20, pp
  735--741, \doi{10.1145/3387940.3392202}

\bibitem[{{Gonzalez-Barahona} et~al.(2017){Gonzalez-Barahona}, Sherwood,
  Robles, and Izquierdo}]{gonzalez-barahonaTechnicalLagSoftware2017}
{Gonzalez-Barahona} JM, Sherwood P, Robles G, Izquierdo D (2017) Technical
  {{Lag}} in {{Software Compilations}}: {{Measuring How Outdated}} a {{Software
  Deployment Is}}. In: Balaguer F, Di~Cosmo R, Garrido A, Kon F, Robles G,
  Zacchiroli S (eds) Open {{Source Systems}}: {{Towards Robust Practices}}, vol
  496, Springer International Publishing, Cham, pp 182--192,
  \doi{10.1007/978-3-319-57735-7_17}

\bibitem[{Guizani et~al.(2023)Guizani, {Castro-Guzman}, Sarma, and
  Steinmacher}]{guizaniRulesEngagementWhy2023}
Guizani M, {Castro-Guzman} AA, Sarma A, Steinmacher I (2023) Rules of
  {{Engagement}}: {{Why}} and {{How Companies Participate}} in {{OSS}}. In:
  Proceedings of the 45th {{International Conference}} on {{Software
  Engineering}}, IEEE Press, Melbourne, Victoria, Australia, {{ICSE}} '23, pp
  2617--2629, \doi{10.1109/ICSE48619.2023.00218}

\bibitem[{Hertzog and Mas(2012)}]{hertzogDebianAdministratorsHandbook2012}
Hertzog R, Mas R (2012) The {{Debian Administrator}}'s {{Handbook}}. Freexian

\bibitem[{Jackson and Schwarz(2024)}]{jacksonDebianPolicyManual2024}
Jackson I, Schwarz C (2024) Debian {{Policy Manual}}.
  https://www.debian.org/doc/debian-policy/ch-controlfields.html\#version

\bibitem[{Jayasuriya et~al.(2023)Jayasuriya, Terragni, Dietrich, Ou, and
  Blincoe}]{jayasuriyaUnderstandingBreakingChanges2023}
Jayasuriya D, Terragni V, Dietrich J, Ou S, Blincoe K (2023) Understanding
  {{Breaking Changes}} in the {{Wild}}. In: Proceedings of the 32nd {{ACM
  SIGSOFT International Symposium}} on {{Software Testing}} and {{Analysis}},
  Association for Computing Machinery, New York, NY, USA, {{ISSTA}} 2023, pp
  1433--1444, \doi{10.1145/3597926.3598147}

\bibitem[{Jayasuriya et~al.(2025)Jayasuriya, Ou, Hegde, Terragni, Dietrich, and
  Blincoe}]{jayasuriyaExtendedStudySyntactic2025c}
Jayasuriya D, Ou S, Hegde S, Terragni V, Dietrich J, Blincoe K (2025) An
  extended study of syntactic breaking changes in the wild. Empirical Software
  Engineering 30(2):42, \doi{10.1007/s10664-024-10563-4}

\bibitem[{Kong et~al.(2024)Kong, Liu, Bao, and
  Lo}]{kongBetterComprehensionBreaking2024}
Kong D, Liu J, Bao L, Lo D (2024) Towards {{Better Comprehension}} of
  {{Breaking Changes}} in the {{NPM Ecosystem}}. ACM Trans Softw Eng Methodol
  \doi{10.1145/3702991}

\bibitem[{Lam et~al.(2020)Lam, Dietrich, and
  Pearce}]{lamPuttingSemanticsSemantic2020}
Lam P, Dietrich J, Pearce DJ (2020) Putting the semantics into semantic
  versioning. In: Proceedings of the 2020 {{ACM SIGPLAN International
  Symposium}} on {{New Ideas}}, {{New Paradigms}}, and {{Reflections}} on
  {{Programming}} and {{Software}}, Association for Computing Machinery, New
  York, NY, USA, Onward! 2020, pp 157--179, \doi{10.1145/3426428.3426922}

\bibitem[{Legay et~al.(2020)Legay, Decan, and
  Mens}]{legayPackageFreshnessLinux2020}
Legay D, Decan A, Mens T (2020) On {{Package Freshness}} in {{Linux
  Distributions}}. In: 2020 {{IEEE International Conference}} on {{Software
  Maintenance}} and {{Evolution}} ({{ICSME}}), pp 682--686,
  \doi{10.1109/ICSME46990.2020.00072}

\bibitem[{Legay et~al.(2021)Legay, Decan, and Mens}]{legay_quantitative_2021}
Legay D, Decan A, Mens T (2021) A {{Quantitative Assessment}} of {{Package
  Freshness}} in {{Linux Distributions}}. In: 2021 {{IEEE}}/{{ACM}} 4th
  {{International Workshop}} on {{Software Health}} in {{Projects}},
  {{Ecosystems}} and {{Communities}} ({{SoHeal}}), pp 9--16,
  \doi{10.1109/SoHeal52568.2021.00008}

\bibitem[{Li et~al.(2022)Li, Moreschini, Pecorelli, and Taibi}]{li_ossara_2022}
Li X, Moreschini S, Pecorelli F, Taibi D (2022) {{OSSARA}}: {{Abandonment Risk
  Assessment}} for {{Embedded Open Source Components}}. IEEE Software
  39(4):48--53, \doi{10.1109/MS.2022.3163011}

\bibitem[{Lin{\aa}ker et~al.(2022)Lin{\aa}ker, Papatheocharous, and
  Olsson}]{linaker_how_2022}
Lin{\aa}ker J, Papatheocharous E, Olsson T (2022) How to characterize the
  health of an {{Open Source Software}} project? {{A}} snowball literature
  review of an emerging practice. In: Proceedings of the 18th {{International
  Symposium}} on {{Open Collaboration}}, Association for Computing Machinery,
  New York, NY, USA, {{OpenSym}} '22, pp 1--12, \doi{10.1145/3555051.3555067}

\bibitem[{Miller et~al.(2023)Miller, K{\"a}stner, and
  Vasilescu}]{miller_we_2023}
Miller C, K{\"a}stner C, Vasilescu B (2023) ``{{We Feel Like We}}'re {{Winging
  It}}:'' {{A Study}} on {{Navigating Open-Source Dependency Abandonment}}. In:
  Proceedings of the 31st {{ACM Joint European Software Engineering
  Conference}} and {{Symposium}} on the {{Foundations}} of {{Software
  Engineering}}, Association for Computing Machinery, New York, NY, USA,
  {{ESEC}}/{{FSE}} 2023, pp 1281--1293, \doi{10.1145/3611643.3616293}

\bibitem[{Nguyen and Holt(2012)}]{nguyenLifeDeathSoftware2012}
Nguyen R, Holt R (2012) Life and death of software packages: An evolutionary
  study of {{Debian}}. In: Proceedings of the 2012 {{Conference}} of the
  {{Center}} for {{Advanced Studies}} on {{Collaborative Research}}, IBM Corp.,
  USA, {{CASCON}} '12, pp 192--204

\bibitem[{Nussbaum and Zacchiroli(2010)}]{nussbaumUltimateDebianDatabase2010}
Nussbaum L, Zacchiroli S (2010) The {{Ultimate Debian Database}}:
  {{Consolidating}} bazaar metadata for {{Quality Assurance}} and data mining.
  In: 2010 7th {{IEEE Working Conference}} on {{Mining Software Repositories}}
  ({{MSR}} 2010), pp 52--61, \doi{10.1109/MSR.2010.5463277}

\bibitem[{Pinckney et~al.(2023)Pinckney, Cassano, Guha, and
  Bell}]{pinckneyLargeScaleAnalysis2023a}
Pinckney D, Cassano F, Guha A, Bell J (2023) A {{Large Scale Analysis}} of
  {{Semantic Versioning}} in {{NPM}}. In: 2023 {{IEEE}}/{{ACM}} 20th
  {{International Conference}} on {{Mining Software Repositories}} ({{MSR}}),
  pp 485--497, \doi{10.1109/MSR59073.2023.00073}

\bibitem[{{Preston-Werner}(2024)}]{preston-werner_semantic_nodate}
{Preston-Werner} T (2024) Semantic {{Versioning}} 2.0.0. https://semver.org/

\bibitem[{RedHat(2014)}]{redhatCVE20140160OpenSSLHeartbleed2014}
RedHat (2014) {{CVE-2014-0160 OpenSSL}} '{{Heartbleed}}' vulnerability.
  https://nvd.nist.gov/vuln/detail/cve-2014-0160

\bibitem[{RedHat(2024)}]{redhatCVE20236246GlibcHeapbased2024}
RedHat (2024) {{CVE-2023-6246}} glibc: Heap-based buffer overflow.
  https://nvd.nist.gov/vuln/detail/CVE-2023-6246

\bibitem[{Stringer et~al.(2020)Stringer, Tahir, Blincoe, and
  Dietrich}]{stringerTechnicalLagDependencies2020}
Stringer J, Tahir A, Blincoe K, Dietrich J (2020) Technical {{Lag}} of
  {{Dependencies}} in {{Major Package Managers}}. In: 2020 27th {{Asia-Pacific
  Software Engineering Conference}} ({{APSEC}}), pp 228--237,
  \doi{10.1109/APSEC51365.2020.00031}

\bibitem[{Tan and Zhou(2019)}]{tanHowCommunicateWhen2019}
Tan X, Zhou M (2019) How to {{Communicate}} when {{Submitting Patches}}: {{An
  Empirical Study}} of the {{Linux Kernel}}. Proc ACM Hum-Comput Interact
  3(CSCW):108:1--108:26, \doi{10.1145/3359210}

\bibitem[{Zerouali et~al.(2018)Zerouali, Constantinou, Mens, Robles, and
  {Gonz{\'a}lez-Barahona}}]{zeroualiEmpiricalAnalysisTechnical2018}
Zerouali A, Constantinou E, Mens T, Robles G, {Gonz{\'a}lez-Barahona} J (2018)
  An {{Empirical Analysis}} of {{Technical Lag}} in npm {{Package
  Dependencies}}. In: Capilla R, Gallina B, Cetina C (eds) New
  {{Opportunities}} for {{Software Reuse}}, vol 10826, Springer International
  Publishing, Cham, pp 95--110, \doi{10.1007/978-3-319-90421-4_6}

\bibitem[{Zerouali et~al.(2019)Zerouali, Mens, {Gonzalez-Barahona}, Decan,
  Constantinou, and Robles}]{zeroualiFormalFrameworkMeasuring2019}
Zerouali A, Mens T, {Gonzalez-Barahona} J, Decan A, Constantinou E, Robles G
  (2019) A formal framework for measuring technical lag in component
  repositories --- and its application to npm. Journal of Software: Evolution
  and Process 31(8):e2157, \doi{10.1002/smr.2157}

\bibitem[{Zerouali et~al.(2021)Zerouali, Mens, Decan, {Gonzalez-Barahona}, and
  Robles}]{zeroualiMultidimensionalAnalysisTechnical2021}
Zerouali A, Mens T, Decan A, {Gonzalez-Barahona} J, Robles G (2021) A
  multi-dimensional analysis of technical lag in {{Debian-based Docker}}
  images. Empirical Software Engineering 26(2):19,
  \doi{10.1007/s10664-020-09908-6}

\end{thebibliography}

\section*{Biography and Photo}

\vspace{1em}

\noindent
\begin{minipage}{0.3\textwidth}
    \includegraphics[width=\textwidth]{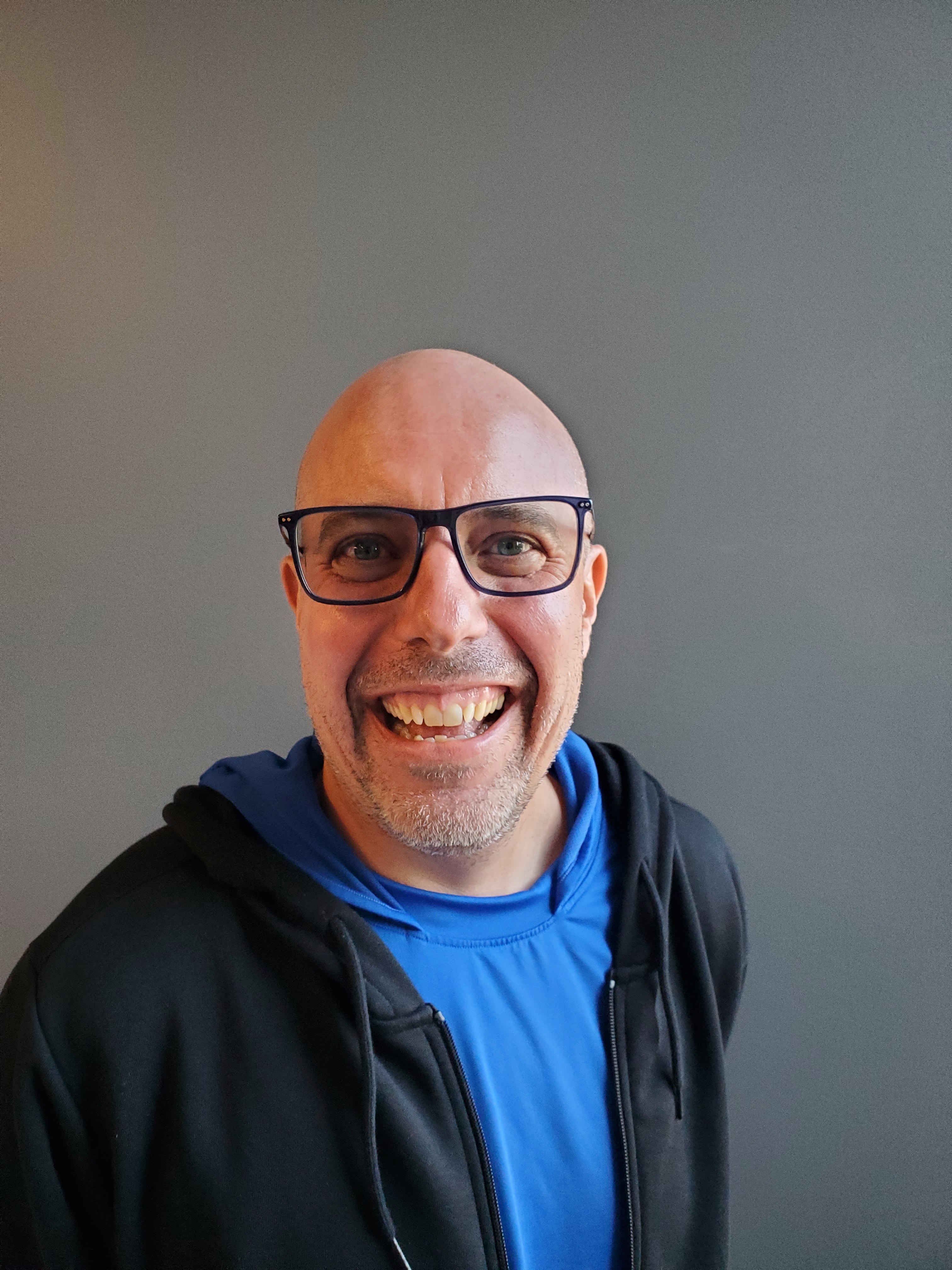} 
\end{minipage}
\hfill
\begin{minipage}{0.65\textwidth}
         \textbf{Shane K. Panter} brings over a decade of industry experience to his role as a Clinical Assistant Professor in the Computer Science Department at Boise State University. His professional background includes extensive work in manufacturing, where he developed and supported large-scale Enterprise Resource Planning (ERP) and Material Requirements Planning (MRP) systems. At Hewlett-Packard, Shane contributed to firmware development, working in kernel and user space to optimize energy efficiency for enterprise laser jet printers. His current research interests focus on empirical software engineering and cybersecurity, particularly within kernel and user space environments. He can be reached at shanepanter@boisestate.edu.
\end{minipage}

\vspace{1em}

\noindent
\begin{minipage}{0.3\textwidth}
    \includegraphics[width=\textwidth]{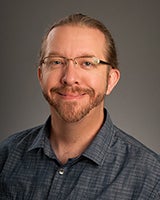} 
\end{minipage}
\hfill
\begin{minipage}{0.65\textwidth}
         \textbf{Lucas S. Hindman} is a Clinical Assistant Professor of Computer Science at Boise State University. His research focuses on neuromorphic computing, high-performance computing, storage systems, and software-defined radio. He holds B.S. and M.S. degrees in Computer Science and is currently pursuing a Ph.D. in Computing with an emphasis in Computer Science at Boise State University, where his doctoral work centers on spiking neural networks. He can be reached at lukehindman@boisestate.edu.
\end{minipage}

\vspace{1em}
\noindent
\begin{minipage}{0.3\textwidth}
    \includegraphics[width=\textwidth]{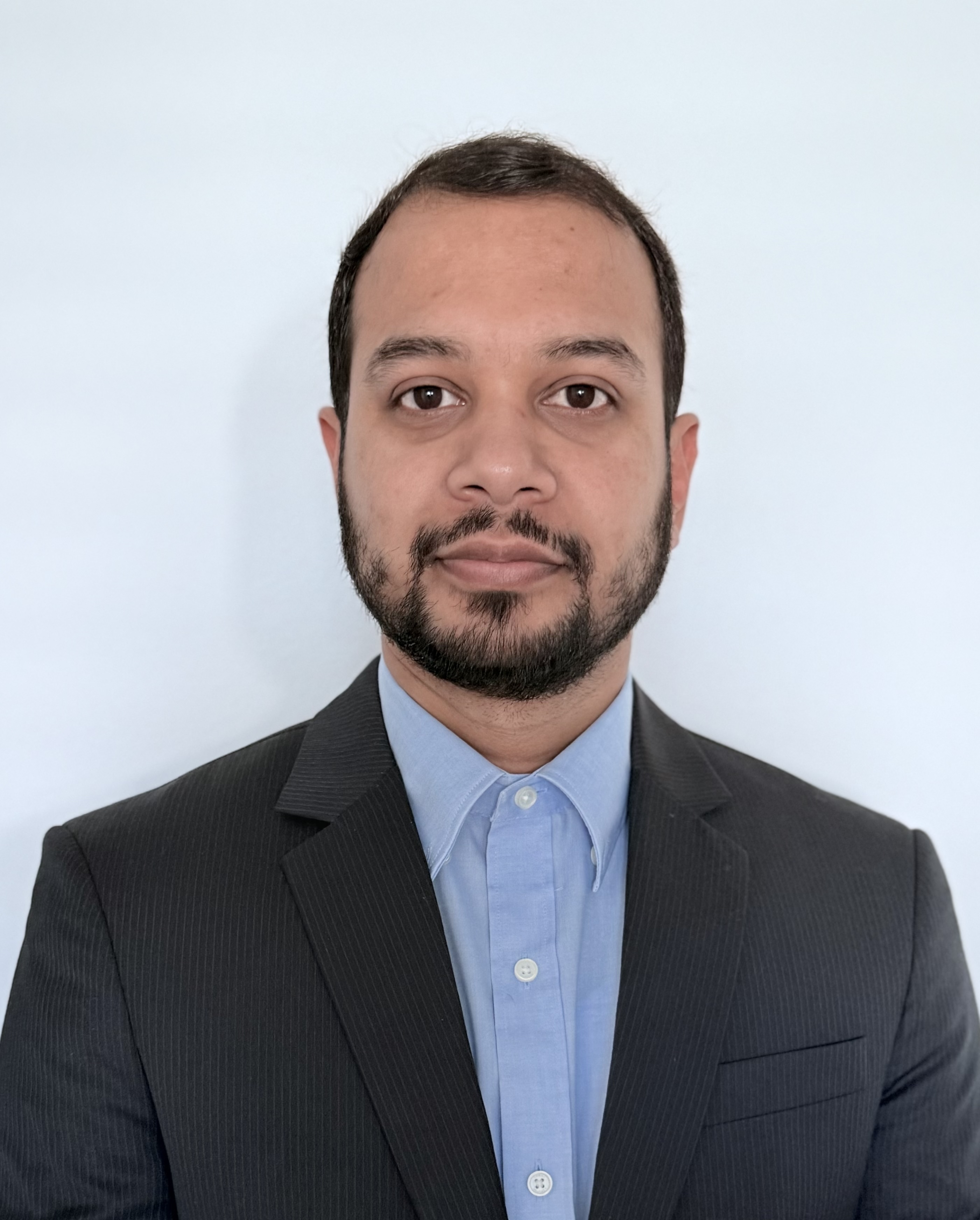} 
\end{minipage}
\hfill
\begin{minipage}{0.65\textwidth}
         \textbf{Nasir U. Eisty} is an Assistant Professor of Computer Science at the University of Tennessee Knoxville. His research interests lie in the areas of Empirical Software Engineering, AI for Software Engineering, Scientific and Research Software Engineering, and Software Security. He received his Ph.D. degree in Computer Science from the University of Alabama. Contact him at neisty@utk.edu
\end{minipage}

\end{document}